\documentclass{article}[12pt]  %

\usepackage{fullpage}
\usepackage{caption}
\usepackage{multirow}
\usepackage{subcaption}
\usepackage{url}
\usepackage{graphicx}
\usepackage{epstopdf}
\graphicspath{{./figures/}}
\usepackage{color}
\usepackage[utf8]{inputenc}

\let\oldenumerate\enumerate
\renewcommand{\enumerate}{
  \oldenumerate
  \setlength{\itemsep}{0pt}
  \setlength{\parskip}{0pt}
  \setlength{\parsep}{0pt}
}

\begin{document}
\newcommand{\narenc}[1]{[{\color{red} Naren writes: \it #1}]}
\newcommand{\patrickc}[1]{[{\color{blue} Patrick writes: \it #1}]}
\newcommand{\sathappanc}[1]{[{\color{green} Sathappan writes: \it #1}]}

\newcommand{\affaddr}[1]{{#1}}

\title{`Beating the news' with EMBERS: \\
Forecasting Civil Unrest using Open Source Indicators}
\author{
  Naren Ramakrishnan\footnotemark[1],
  Patrick Butler\footnotemark[1],
  Sathappan Muthiah\footnotemark[1],
  Nathan Self\footnotemark[1], \\
  Rupinder Khandpur\footnotemark[1],
  Parang Saraf\footnotemark[1],
  Wei Wang\footnotemark[1],
  Jose Cadena\footnotemark[1],
  Anil Vullikanti\footnotemark[1],\\
  Gizem Korkmaz\footnotemark[1],
  Chris Kuhlman\footnotemark[1],
  Achla Marathe\footnotemark[1],
  Liang Zhao\footnotemark[1],
  Ting Hua\footnotemark[1],
  Feng Chen\footnotemark[8],\\\
  Chang-Tien Lu\footnotemark[1],
  Bert Huang\footnotemark[2],
  Aravind Srinivasan\footnotemark[2],
  Khoa Trinh\footnotemark[2],
  Lise Getoor\footnotemark[3],\\
  Graham Katz\footnotemark[4],
  Andy Doyle\footnotemark[4],
  Chris Ackermann\footnotemark[4],
  Ilya Zavorin\footnotemark[4],
  Jim Ford\footnotemark[4],\\
  Kristen Summers\footnotemark[4],
  Youssef Fayed\footnotemark[5],
  Jaime Arredondo\footnotemark[6],
  Dipak Gupta\footnotemark[7],
  David Mares\footnotemark[6]
\\
\affaddr{\footnotemark[1]~~Virginia Tech, Blacksburg, VA 24061}\\
\affaddr{\footnotemark[2]~~University of Maryland, College Park, MD 20742}\\
\affaddr{\footnotemark[3]~~University of California at Santa Cruz, Santa Cruz, CA 95064}\\
\affaddr{\footnotemark[8]~~~University at Albany - State University of New York, Albany, NY 12222}\\
\affaddr{\footnotemark[6]~~University of California at San Diego, San Diego, CA 92093}\\
\affaddr{\footnotemark[7]~~~San Diego State University, San Diego, CA 92182} \\
\affaddr{\footnotemark[5]~~BASIS Technology, Herndon, VA 20171}\\
\affaddr{\footnotemark[4]~~CACI Inc., Lanham, MD 20706}\\
Contact: \{pabutler@vt.edu, naren@cs.vt.edu\}\\
}

\maketitle
\begin{abstract}
We describe the design, implementation, and evaluation of EMBERS, an automated, 24x7
continuous system for forecasting civil unrest across 10 countries of
Latin America using open source indicators such as tweets, news sources, blogs, economic indicators, and
other data sources. Unlike retrospective studies, EMBERS has been making forecasts into the future
since Nov 2012 which have been (and continue to be) evaluated by an independent
T\&E team (MITRE). 
Of note, EMBERS has successfully forecast the uptick and downtick of incidents during
the June 2013 protests in Brazil. 
We outline the system
architecture of EMBERS, individual models that leverage specific data sources, and
a fusion and suppression engine that supports trading off specific evaluation
criteria. EMBERS also provides an audit trail interface that enables the investigation of
why specific predictions were made along with the data utilized for forecasting. Through
numerous evaluations, we demonstrate the superiority of EMBERS over baserate methods
and its capability to forecast significant societal happenings.

\end{abstract}

\vspace{-1em}
\section{Introduction}
We are constantly reminded of instabilities across the world, e.g., in regions such as Middle East and
Latin America. Some of these instabilities arise from extremism or terrorism while others are
the result of civil unrest, involving population-level uprisings by disenchanted citizens.
Since the Arab Spring revolution began, and especially after Egypt's upheaval, many analysts
(e.g., \cite{Leetaru2011}) 
have pondered: Could we have anticipated these events? Were there precursors and signals that could
have alerted us to them? Why did this happen in one country but not another?

Our team is an industry-university partnership charged with developing a system to continually monitor
data sources 24x7, mine them
to yield emerging trends, and process these trends into forecasts of significant societal events such as protests.
We refer to our system as EMBERS for Early Model Based Event Recognition using Surrogates.
Although the scope of EMBERS spans a broad class of events (e.g., protests, disease outbreaks, elections), we
focus our attention in this paper on only civil unrest events.
Civil unrest is defined as a population-level
event wherein people protest against the government or other larger organizations
about specific policies, issues, or situations.

The EMBERS project is supported by the IARPA (Intelligence Advanced Research Projects Activity) OSI (Open Source
Indicators) program whose objective is to forecast population-level changes using open source data feeds, such as tweets, web searches, news/blogs,
economic indicators, Wikipedia, Internet traffic, and other sources.
(The term `open source' here refers to data sources that are openly available without requiring
privileged access.)
As a performer in the OSI program, EMBERS is a deployed system that has
been generating forecasts since Nov 2012 and automatically emailing them in real-time
to IARPA upon generation, which have been evaluated
by an independent test and evaluation (T\&E) team (MITRE). Using human analysts,
MITRE organizes a gold standard report (GSR) of protests 
by surveying newspapers for reportings of civil unrest.
Our forecasts have been evaluated against this GSR every month since Nov 2012.
Thus, unlike studies of retrospective
predictability, EMBERS has been generating (and continues to generate) forecasts into the future.

\begin{figure}[t]
\centering
\includegraphics[scale=0.2,width=0.50\textwidth,height=0.2\textheight]{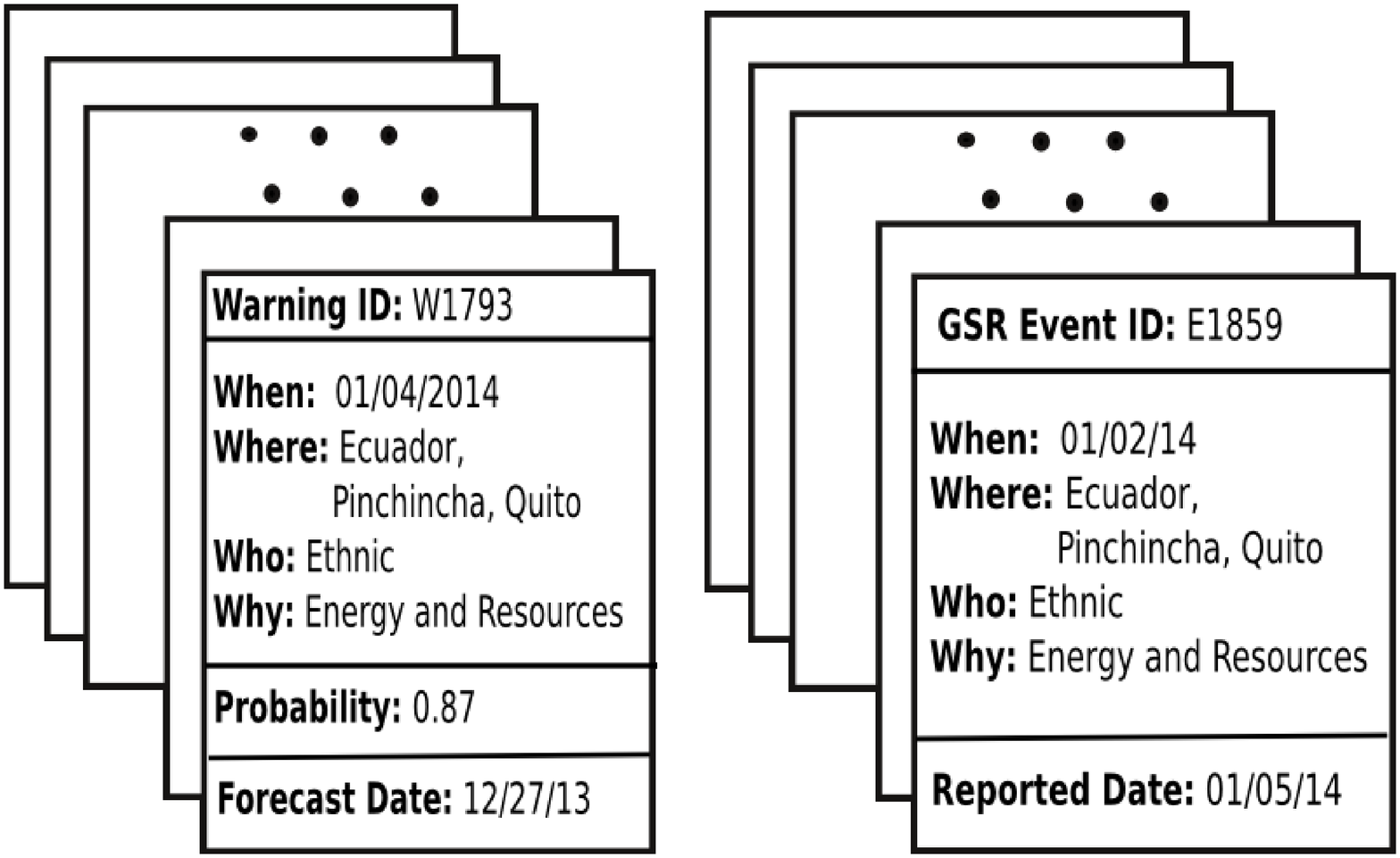}
\caption{Alerts (left) and events (right) are
structured records describing protests.}
\label{fig:alertstructure}
\end{figure}

Our goal in this paper is to present the design, implementation, and evaluation of EMBERS over an extended
period of time.
Our key contributions are:
\begin{enumerate}
\item We outline the system architecture and design of EMBERS, a modular `big data' processing environment with
levels of data transduction from raw feeds to warnings. EMBERS's
alerts are meant for analyst consumption, but
the system runs continuously 24x7 {\it without} a human-in-the-loop.
\item Unlike other forecasting/warning generation systems with
similar motivations (e.g.,~\cite{icews}),
EMBERS warnings are highly structured, capturing (i) when the protest is forecast to happen,
(ii) where, with a city-level granularity, (iii) which subgroup of the population will protest,
(iv) why will they be protesting, and (v) a probability associated with the forecast. See Fig.~\ref{fig:alertstructure} (left)
for an example of what an alert looks like.
\item EMBERS adopts a multi-model approach wherein each model harnesses different data sources to independently
generate predictions and such predictions are then fused to yield final warnings. 
Using the formalism of probabilistic soft logic (PSL~\cite{broecheler:uai10}), we demonstrate how we
can leverage the selective superiorities of different models and how we can employ
collective reasoning to
help `shape' predictions into a final set of warnings.
\item We illustrate the application of EMBERS to 10 countries in Latin America, viz. 
Argentina, Brazil, Chile, Colombia, Ecuador, El Salvador, Mexico, Paraguay, Uruguay, and Venezuela.
We present an exhaustive suite of experiments evaluating EMBERS 
w.r.t. multiple forecasting criteria and for its capability to forecast significant 
societal events
such as the June 2013 protests in Brazil, also known as the `Brazilian Spring.' \end{enumerate}

\vspace{-1em}
\section{What is Civil Unrest?}
Event analysis of the form considered here
is an established concept in social science
research~\cite{Box-Steffensmeier2004}.  
Civil unrest is a large concept intended to capture the myriad ways in which
people express their  protest against things that affect their lives and for
which they assume that the government (local, regional or national) has a
responsibility  (e.g., cost of urban transportation, poor infrastructure,
etc.).  If the action is directed against private actors, there is normally a
connection to government policy or behavior, e.g., a labor strike against a
private company can disrupt the rhythm of everyday life for the rest of
society, turn violent or lead to a series of disruptive strikes which require
government involvement, and thus responsibility in the eyes of citizens. Civil
unrest does not include acts by criminals for purely private gain. While
authoritarian governments may outlaw civil protest and thus `criminalize' the
participants, social scientists would distinguish illegal political protests
from illegal criminal activities. Gang members stopping public buses to extort
payoffs from bus owners would not be a civil unrest event, though people
protesting afterward against the government's inability to control such gangs
would be considered civil unrest.

This expansive definition of civil unrest means that one can find it
everywhere, including European protests against austerity or marches against an
oil pipeline from Canada across the US to the Gulf of Mexico. Latin America,
nevertheless, offers some special characteristics that make it an excellent
region for study in our project.
The region experiences a plethora of civil unrest events every day (providing a
sufficient number of GSR events to train machine learning models), is well covered by international and
national news media (facilitating the task of generating ground truth), is the
object of detailed empirical research and polling (permitting the description
of the social, political and economic context within which civil unrest occurs)
and has a significant and growing number of social network users (thus supporting
the use of modern data mining algorithms).

\vspace{-1em}
\section{Related Work}
Three broad classes of related work pertinent to EMBERS, are briefly surveyed here.
First, there is a rich body of literature in {\bf event coding}~\cite{philschrodt,texasguy}
wherein structured descriptions of events are extracted from text (e.g., news reports). ICEWS~\cite{icews} and GDELT~\cite{Leetaru2013}
are two prominent systems for event coding and significant work has built upon them to develop
predictive systems. For instance, ICEWS-coded events have been utilized to forecast the possibility of
domestic political crises within countries.
Second, there is considerable research on {\bf civil unrest modeling}
although much of this work focuses on characterization rather than forecasting.
The dynamics by which volunteers are recruited via social networks
to the May 2011 Spain protests
was studied in~\cite{sandra}. Spatial and temporal
distributions of civil unrest over 170 countries were studied in~\cite{braha}.
There are many papers that aim to retrospectively analyze the breadcrumbs of information
preceding significant events such as the Arab Spring~\cite{Leetaru2011, newpaperbyJason}.
Our group has analyzed protests in Latin America paying specific attention to signals that manifest in
social media~\cite{tingpaper}. Finally, there is a growing body of work
on {\bf event extraction}. Companies like RecordedFuture and works like~\cite{gravano,etzioni}
aim to recognize planned (future) events from text and Twitter. EMBERS distinguishes from all of the
above by supporting highly structured descriptions of protests,
emphasizing forecasting rather than characterization or mere detection,
and utilizing a broader range of data sources than prior work. Finally, we reiterate that EMBERS is
a deployed system that has been successfully issuing alerts for the past 15 months.

\vspace{-1em}
\section{System Architecture}

The EMBERS system is a modular, data-analytics platform for generating
warnings of the form described in Fig.~\ref{fig:alertstructure} (left).
It continuously monitors streams
of open source data and generates structured alerts in real time,
delivered by
email to IARPA/MITRE for scoring, with the date of
email delivery being the {\em forecast date}.

\begin{figure}[t]
\centering
\includegraphics[width=0.30\textwidth]{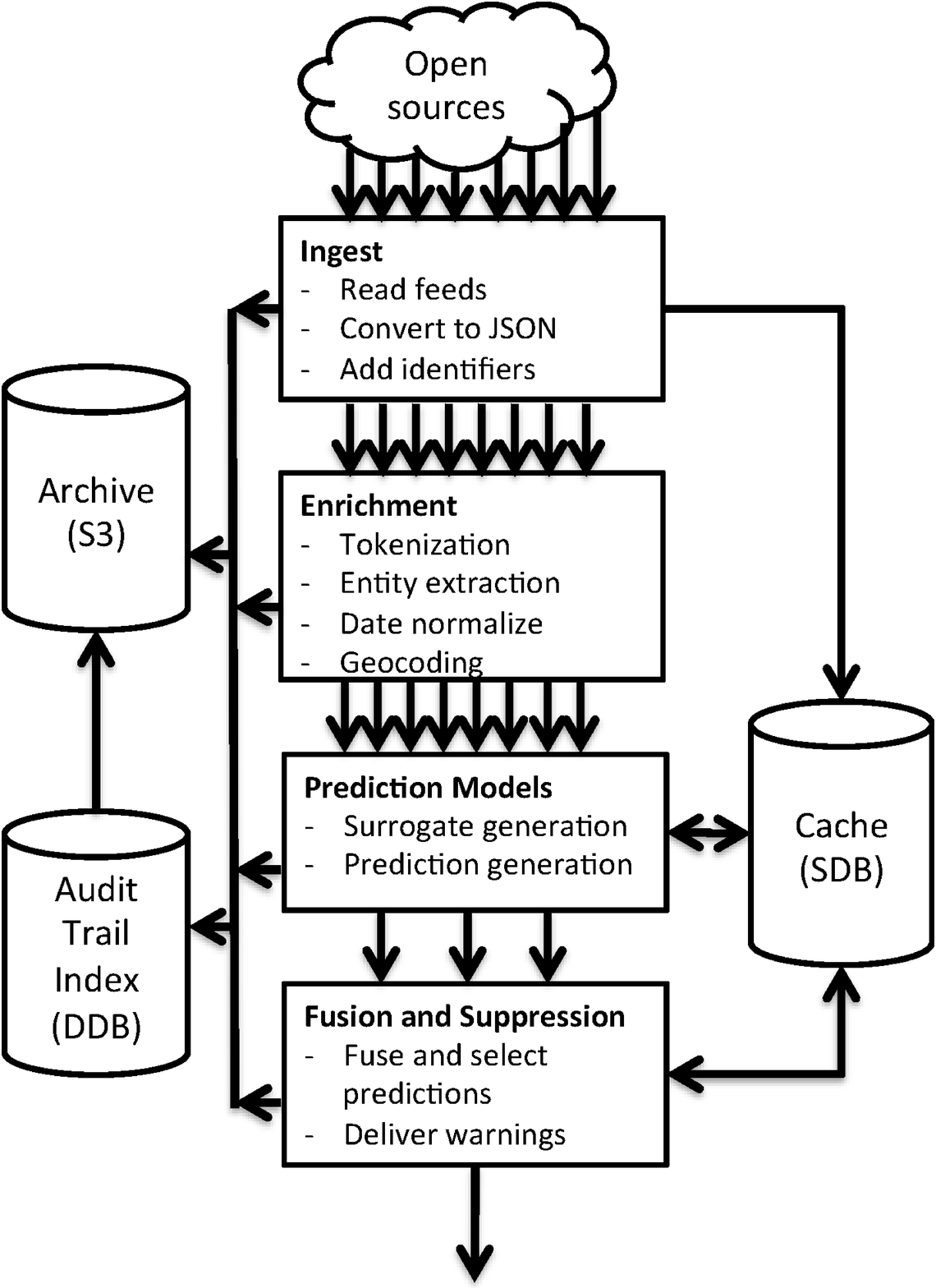}
\caption{EMBERS system architecture}
\label{fig:arch}
\end{figure}

The EMBERS architecture, illustrated in Fig.~\ref{fig:arch}, provides a
platform for the ingest and warehousing of a variety of raw data sources,
and a flexible mechanism for data transfer among ingest, analytics
and prediction modules.  The four stages---ingest, enrichment, prediction, delivery---are
described in detail, respectively, in Sections~\ref{ingest},~\ref{enrich},~\ref{model}, and~\ref{fuse}.
EMBERS runs in the commercial AWS cloud. It
implements a share-nothing, message-based, streaming architecture
using 0MQ as the underlying method of data transport.  Processing
components are distributed among virtual machines in a configurable,
network-secure, auto-deployable, cluster of EC2 instances. With
loosely coupled processes and configuration driven communication,
EMBERS is able to deliver warnings reliably while facilitating rapid
integration and deployment of new components and data sources.
The current production cluster consists of 12 EC2 instances with two
dedicated to ingest processing, three dedicated to message enrichment,
four dedicated to predictive modeling and warnings selection, one each dedicated
to archiving and system monitoring. EMBERS became operational in November 2012. It has
ingested nearly 13TB of raw data and generated over 12,000 warnings
since then. Other notable statistics are listed in
Table~\ref{tab:stats}.

\begin{table}
\caption{EMBERS system statistics}
 \centering
 \begin{tabular}{|l|l|l|l|l|}
 \hline
 Archived data     & 12.4 TB                  \\ \hline
 Archive size & ca. 3 billion messages   \\ \hline
 Data throughput   & 200-2000 messages/sec  \\ \hline
 Daily ingest & 15 GB \\ \hline
 System memory & 50 GB \\ \hline
 System core & 16 vCPUs \\ \hline
 System output & ca. 40 warnings/day \\ \hline
\end{tabular}
\label{tab:stats}
\end{table}

\subsection{Ingest Processing}
\label{ingest}
The EMBERS ingest module processes data from a variety of different
sources: Twitter's public API, Datasift's processed Twitter feed,
Healthmap's alerts and reports, RSS news and blog feeds, Talkwalker
alerts, NASA satellite meteorological data, Google Flu Trends, Bloomberg
financial news, TOR usage data, OpenTable's restaurant cancellation data, the
PAHO health survey, and web-pages referenced Tweets. (Some of these, e.g.,
NASA satellite data and Google Flu Trends are used for other event classes,
as described in the introduction.)  Each of these has
a dedicated configurable ingest processor.  Ingested data is packaged
into UTF8-encoded JSON messages, assigned unique identifiers and
published to a source-specific queue, allowing for simple archiving
and subscription.  Simple time-series and systems data, such as the store of warnings sent,
are stored in a database cache.

One of our central ingest processes makes use of Datasift's Twitter
collection engine. Datasift provides the ability to query and stream
tweets in real time.  These tweets are augmented with various types of
metadata including the user profile of the tweeting user or geotagged
attributes and the query can target any of these.  Targeting tweets
that come from a particular geographic area, e.g. Latin America, can
be tricky.  While some tweets use geotags to specify the location of
the tweet, these tweets only comprise about $5\%$ of the total number
of tweets and may not be representative of the population overall
(i.e. geotagged tweets come
from people who have smart phones who also tend to be more affluent).
Therefore, it is important to use other information to build a query
that targets relevant tweets.  In building our query we consider
geotag bounding boxes (structured geographical coordinates), Twitter
Places (structured data), user profile location (unstructured,
unverified strings), and finally mentions of a location contained in
the body of the tweet.

\subsection{Message Enrichment}
\label{enrich}

\begin{figure}
  \begin{center}
    \includegraphics[width=\columnwidth]{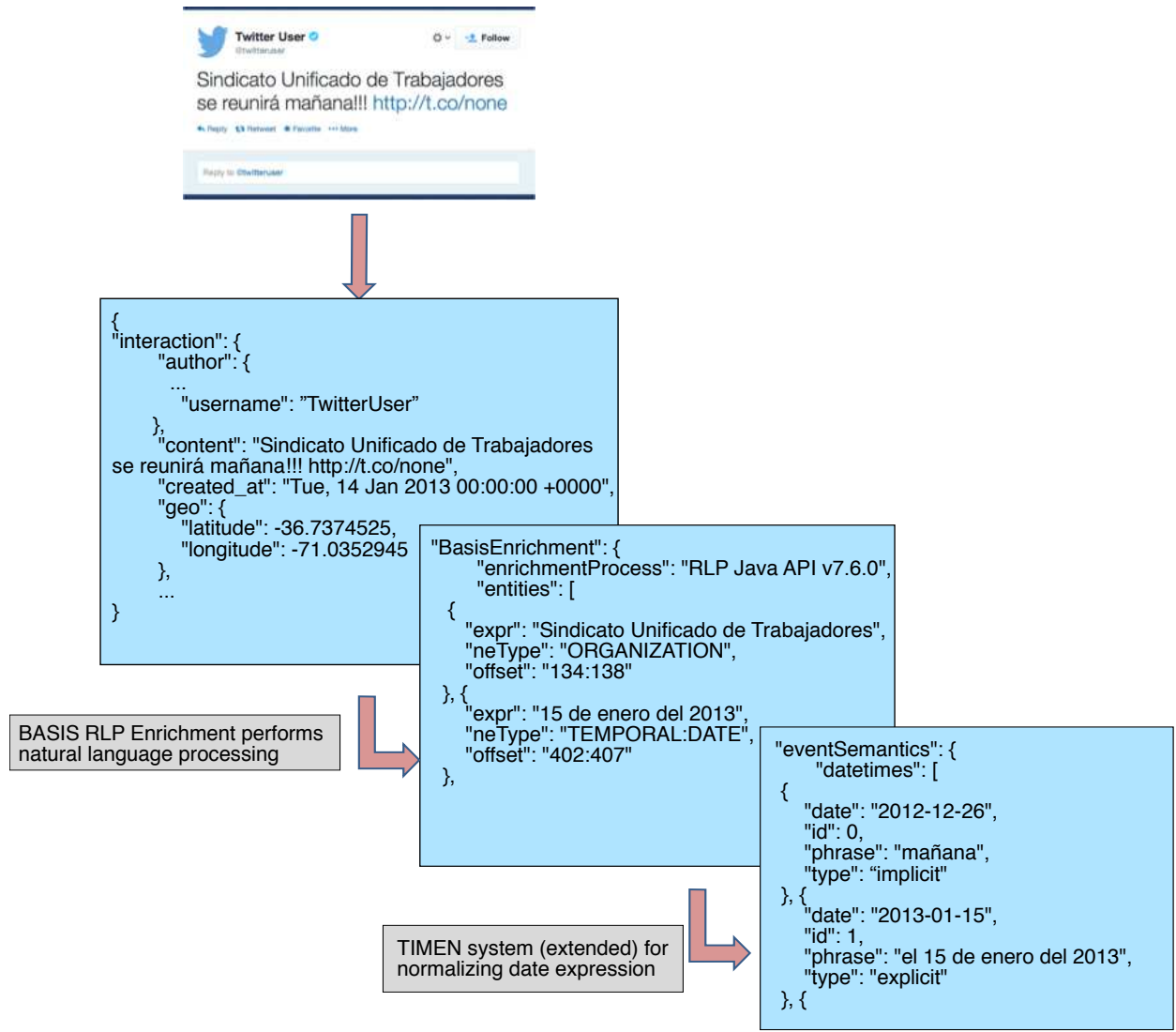}
  \end{center}
  \caption{The process of enriching a tweet using Basis RLP enrichment and TIMEN enrichment to generate exact dates.
    The phrase
    ``Sindicato Unificado de Trabajadores se reunir\`a mana\~na''
    gets enriched to
    ``Sindicato Unificado de Trabajadores se reunir\`a January 15, 2013.''
  }
  \label{fig:enrichment}
\end{figure}

Messages with textual content (tweets, newsfeeds, blog postings, etc.)
are subjected to shallow linguistic processing prior to analysis. Note that
most of our content involves languages from the Latin American region, esp.
Spanish, Portuguese, but also French (and of course, English). Applying
BASIS technologies' Rosette Language Processing (RLP) tools, the language of
the text is identified, the natural language content is tokenized and
lemmatized and the named entities identified and classified. Date
expressions are normalized and deindexed (using the
TIMEN \cite{LlorensDGS12} package).  Finally, messages are geocoded with a specification
of the location (city, state, country), being talked about in the message.
An example of this enrichment processing can be seen in
Fig.~\ref{fig:enrichment}.

The EMBERS system makes use of two geocoding systems, one for Tweets
and one for news and blog articles. The Twitter geolocater
determines not only the city, country and state, but also the
approximate latitude and longitude coordinates that a tweet is
referring to, or coming from.  Geocoding is achieved by first
considering the most reliable but least available source,
viz. geotags, which give us exact geographic locations that can be
reverse geocoded into place names.  Second, we consider Twitter places
and use place names present in these fields to geocode the place names
into geographical coordinates.  Finally, we consider the text fields
contained in the user profile (location, description) as well as the
tweet text itself to find mentions of relevant locations which can
then be geocoded into geographical coordinates.

Most news articles and blog posts mention multiple locations, e.g.,
the location of reporting, the location of the incident, and locations corresponding
to the hometown of the newspaper. We developed a probabilistic reasoning
engine using probabilistic soft logic (PSL~\cite{broecheler:uai10})
to infer the most
likely city, state,
and country which is the main geographic focus the article.  The PSL
geocoder combines various types of evidence, such as named entities
such as locations, persons, and organizations identified by RLP, as
well as common names and aliases and populations of known
locations. These diverse types of evidence are used in weighted rules
that prioritize their influence on the PSL model's location
prediction. For example, extracted location tokens are strong
indicators of the content location of an article, while organization
and person names containing location names are weaker but still
informative signals; the rules corresponding to these evidence types
are weighted accordingly.

%
%
%
%
%
%
%
%
%
%

%
%
%
%

\vspace{-1em}
\section{Prediction Models}
\label{model}
We now outline the five different models considered in our study (see Table~\ref{tab:models}), 
paying specific attention to their underlying assumptions, data sources, and
scenarios of applicability. 

\begin{table}
\caption{The five different prediction models in EMBERS.}
 \centering
 \begin{tabular}{|l|l|}
 \hline
{\bf Model} & {\bf Data sources} \\ \hline
Planned protest & RSS (news, blogs), Tweets, \\
& Facebook  \\ \hline
Volume-based & RSS (news, blogs), Tweets, Exchange\\
& rates, TOR, ICEWS, GDELT\\ \hline
DQE & Tweets \\ \hline
Cascades & Tweets \\ \hline
Baseline & GSR \\ \hline
\end{tabular}
\label{tab:models}
\end{table}

\subsection{Planned Protest}
\newcommand{\grahamc}[1]{[{\color{blue} graham writes: \it #1}]}
Many civil unrest events are planned and organized through calls-for-action
by opinion and community leaders who galvanize support for their case.
The planned protest model aims
at detecting such civil unrest events from
traditional media (e.g., news pages, mailing
lists, blogs) and from social media (e.g., Twitter, Facebook).
The model filters the input streams by matching to a custom
multi-lingual lexicon of expressions such as {\em preparación huelga},
{\em llamó a acudir a dicha movilización} or {\em plan to strike}
which are likely to indicate a planned unrest event.  The phrase
matching is done in flexible manner making use of the lemmatized,
tokenized output of the BASIS enrichment module, to allow for variation and
approximations in the matching.  Messages that match are then screened
for the mention of a future time/date occurring in the same sentence as the
phrase. The event type and population are forecast using a multinomial naive
Bayes classifier. Location information is determined using the enrichment geocoders.
The phrase dictionary is thus a crucial aspect of the planned protest
model and was populated in a semi-automatic manner using
both expert knowledge and a simple bootstrapping methodology.

The planned protest model reads three kinds of input messages:
standard natural language text (RSS news and blog feeds, as well
as the content of web pages mentioned in tweets), microblogging text
(Twitter), and Facebook Events pages.
The RSS feeds and web pages are processed as discussed above. For
tweets, in addition to the above processing, we require that the tweet
under consideration be retweeted a minimum number of
times, to avoid erroneous alerts. (This value is set to 20
in our system.) For Facebook, we use their public API to search
for event pages containing the word protest or its synonyms.
Most such Facebook event pages already
provide significant information such as the
planned date of protest, location (sometimes with resolution up to
street level), and population/category of
people involved.

\subsection{Volume-based Model}
Next, we developed a traditional machine learning model
to map from a large set of volume-based
features to protest characteristics.
We use a logistic regression model with LASSO 
(Least Absolute Shrinkage and Selection Operator~\cite{Tibshirani1996})
to select a sparse feature set,
and to predict the probability of occurrence of civil unrest events in
different countries. Tweets are one of the primary inputs to this model.
Country-level tweets are first filtered using a keyword dictionary which includes 614
civil unrest related words (such as protest, riot), 192 phrases (e.g., right to
work), and country-specific actors (public figures, political parties, etc.).
For each keyword, its translations in Spanish, Portuguese and English are also
used for filtering. In order to reduce the noise in the data, only tweets
containing at least 3 keywords are considered. The covariates in the LASSO
regression include (i) daily counts of these protest related keywords in
filtered tweets, (ii) daily counts of the same keywords in news and blogs,
(iii) the exchange rate (country specific currency against dollar), (iv) count
of requests to TOR, i.e., the number of online users who have chosen
to conceal their location and identity from the online community, (v) count of
ICEWS events i.e. events identified by the ``Integrated Conflict Early Warning
System''~\cite{icews}, (vi) average intensity of the ICEWS events, (vii) the
counts of events in publicly available GDELT (Global Data on Events, Location
and Tone) dataset~\cite{Leetaru2013}, which is a record of events in
the international system over multiple decades, and (viii) the average tone and
the Goldstein scale of these events.  A threshold for the probability is
determined by maximizing the area under the ROC. This methodology allows for
detection as well as prediction of country-specific civil unrest events.

\subsection{Dynamic query expansion (DQE)}
\begin{figure*}[tb]
  \begin{center}
    \includegraphics[width=\textwidth]{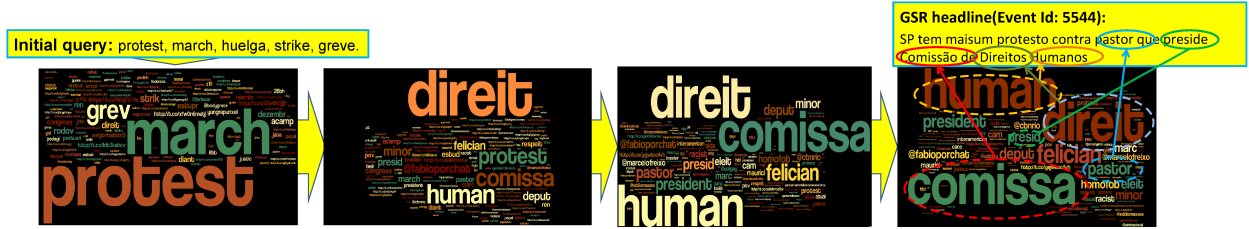}
  \end{center}
  \caption{Steps to building a vocabulary using the DQE model. Beginning from a few general phrases about protests, DQE
hones in on keywords related to a specific GSR event.}
  \label{fig:dqe}
\end{figure*}

The dynamic query expansion (DQE) model is based on the idea that the causes for protests
can be quite varied and, unlike the Volume model (which uses a fixed set of keywords), we must seek emerging conditions for
protests by dynamically growing our vocabularies of interest. This model relies exclusively on tweets.
Given a short seed query, DQE first adopts an iterative keyword expansion strategy to dynamically generate a set of 
extended keywords and tweets pertinent to such keywords. In particular, the seed query consists of a small set of
civil unrest related keywords like ``protest" and ``march.' 
In the initial iteration, we extract the tweets matching the seed query, and rank the terms in them by their DFIDF weights.
Higher ranked terms are used to trigger the second iteration, continuing the process. The iterations are terminated once the set of
keywords and their weights become stable (we have observed that DQE converges in approximately 3--5 iterations). See Fig.~\ref{fig:dqe}.
The resulting tweets are clustered using local modularity and spatial scan statistics, and tweets in the discovered clusters are
used by a classification engine to trigger an alert and to determine the event type and population.

\subsection{Cascades Model}
The cascades model is specifically designed to track activity on
social media, especially recruitment of individuals to causes through the use of 
targeted campaigns, or the popularization of causes through adoption of
hashtags. We characterize information diffusion on a (directed) Twitter
network using
activity cascades. An activity cascade is defined in the following manner: a
user posts a tweet; if one of the followers of this user also posts a tweet on
the same general topic within a short interval of time after the original
poster, we say that the second user was influenced by the first one, and we add
this second tweet to the cascade. Then, we consider the followers of the second
user, and add them to the cascade if they post a tweet within a short interval of
time from then, and so on. The cascade stops growing when none of the
followers of the users in the cascade tweet in the general topic soon enough.
In our model, we compute cascades over two different networks: the follower
graph, which indicates who follows whom in Twitter, and the mention-retweet
graph, where the out-neighbors of a user are those who mention or retweet that
user. Activity cascades are computed for each day (which potentially could have originated
from earlier days and continued growing) and their structural properties
(e.g., size, number of participants, duration) are used as input to a machine learning
model (generalized linear model; GLM) to forecast the probability of occurrence of a GSR event in the
same topic on the following day.
\subsection{Baseline model}
We also developed
a maximum likelihood estimate (MLE) baseline model,  making heavy use of the
GSR. The idea behind this
model is that, even in absence of any explicit signal, the distribution of
events that have appeared in the recent past is a good guide to those
civil unrest events that will take place in the future.
The baseline model makes predictions on the basis of the
distribution of ``event schema''-frequency in the most recent part of
the GSR. An event schema is a combination of a location, an event
type, a population and a day of the week. Some high-frequency
schemas can appear as many as
10 times in a three-month window, but the vast majority of event schemas
appear only once. In a typical three month interval two thirds appear
once with the remaining third split evenly between those
that appear twice, and those that appear three or more times.
Warnings are generated 
with a minimum threshold of 2
and a three-month training interval, and issued 
with a lead time of two weeks.

\vspace{-1em}
\section{Fusion and Suppression}
\label{fuse}
The fusion and suppression engine is responsible for the generation
of the final set of warnings to be delivered. It performs
several key operations:

\begin{itemize}
    \item \textbf{Duplicate detection and warning updating:} Because our prediction models share data sources and the hypothesis space,
duplicate detection is compulsory.
An alert is declared as a duplicate (and discarded) if it shares the same $\langle$ location, event type, population, eventDate $\rangle$ tuple  as a previously
issued alert.
If two alerts differ in only the predicted event dates and those dates are at most
2 days apart, then the alerts are considered to be the same event and an update is issued to the already issued alert.
    \item \textbf{Filling missing values:} Certain models are incapable of predicting all details of an alert such as event type, population, or location up to the city level.
In such cases, the missing information is filled in based on the likelihood of their appearance in the GSR.
    \item \textbf{Warning rewriting:} At times, a model produces a warning with an improbable $\langle$ location, event type, population $\rangle$ combination. Such
a prediction, could either be (1) true, (2) a result of noisy data, or (3) some inherent model error. If the last possibility, one can assume that
the model would have identified the broader region correctly. Under such conditions, the fusion model aims to re-write the predicted city to a
city that is historically most probable within a given radius, and fills in other aspects accordingly.
    \item \textbf{Balancing the recall-quality tradeoff:} It is desirable to sacrifice some amount of recall if our overall objective
is to achieve higher quality of warnings (defined in detail in the next section).  We developed two classes of models to
explore this tradeoff. First, we developed a random forest regression model to predict
likely quality of an alert and alerts that do not pass a desired threshold are suppressed. Second, we trained a
PSL engine on matched alerts and events, to learn probabilities of suppressing warnings based on characteristics of the event predicted.
We explore the performance of both mechanisms in our results.
\end{itemize}

\section{Audit Trail Interface}
\begin{figure}
  \begin{center}
    \includegraphics[width=\columnwidth]{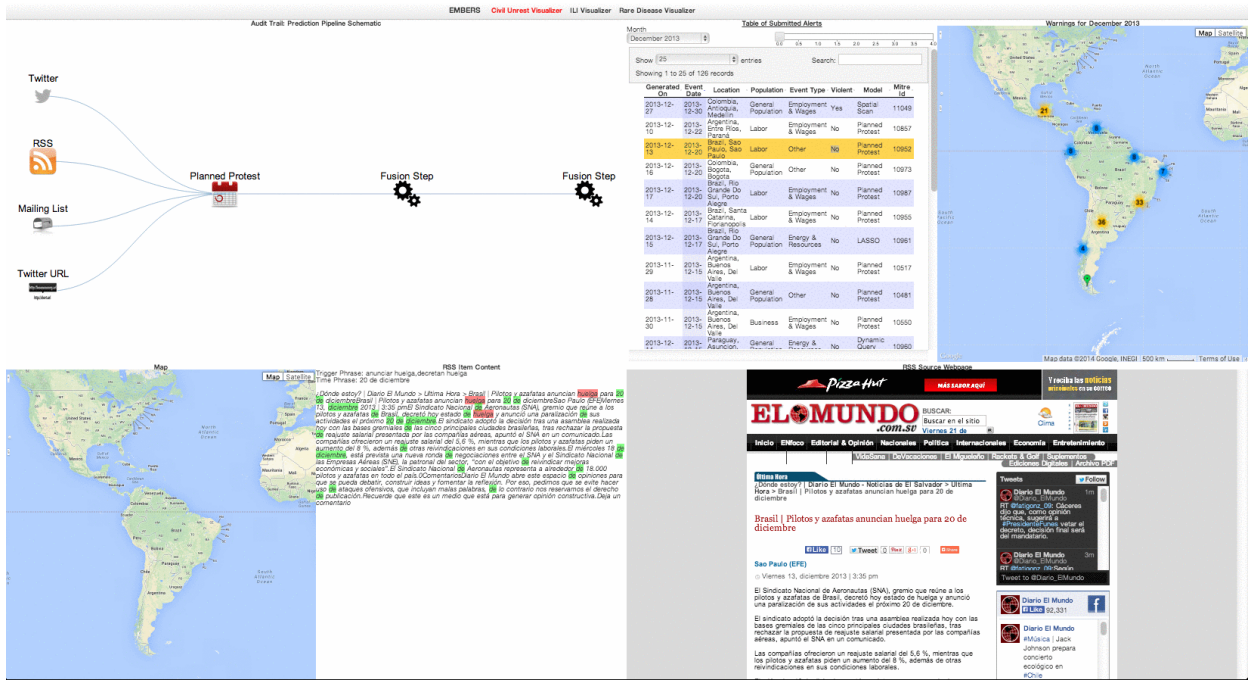}
  \end{center}
  \caption{The audit trail visualization interface, displaying the audit trail
  for an alert from the planned protest model. Explore it at {\bf http://embers.cs.vt.edu\hskip0ex /embers\hskip0ex /alerts.} (top left) Schema of the
planned protest model. (top right) Alert chooser. (bottom panels) Data sources used in this alert, including
highlighted sections.}
  \label{fig:audittrail}
\end{figure}

In order to facilitate auditing of the warnings and further training of
the models, all data that flows through the system is archived to the
Amazon S3 cloud and the processing chain recorded in a NoSQL
database. Using this infrastructure, the EMBERS system can produce an audit
trail for any warning generated, which specifies completely which
messages and analytic processes led to the warning. This audit trail
can be visualized using the EMBERS web-based dashboard, shown in Fig.~\ref{fig:audittrail}.
The interface enables an analyst to rapidly search through warnings, identify the models (and post-processing)
that gave rise to an alert, and the individual data sources that contributed to the alert.

\vspace{-1em}
\section{Evaluation Methodology}
Before we describe our evaluation metrics, it is helpful to review the
composition of alerts and GSR events.
As introduced in Fig.~\ref{fig:alertstructure} (left), an alert
is a structured record containing four
aspects:
(i) the where/why/when/\hskip0ex who of the protest,
(ii) confidence associated with the forecast,
and (iii) (implicitly)
the date the forecast is being made ({\it forecast date}).
The `when' is specified in granularities of days.
The {\bf where} provides a tiered description specifying the
(country, state, city), e.g.,
(Honduras, Francisco Morazan, Tegucigalpa).
The {\bf why} (or event type)
captures the main objective or reason for a civil unrest event,
and is meant to come from 7 broad classes (e.g., `Employment \& Wages',
`Housing', `Energy \& Resources' etc.) each of which is further categorized into
whether the event is forecast to be violent or not.
Finally, the {\bf who} (or population)
denotes common categories of human populations
used in event coding~\cite{philschrodt}
such as
Business, Ethnic, Legal (e.g. judges or lawyers), Education (e.g. teachers or students or parents of students), Religious (e.g. clergy), Medical (e.g., doctors or nurses), Media, Labor, Refugees/Displaced, Agricultural (e.g. farmers,
or just General Population.

Concomitant with the definitions in the above section, a GSR event contains
again the where/why/when/\hskip0ex who of a protest that has actually occurred and
a {\it reported date} (the date a newspaper reports the protest as
having happened).
See Fig.~\ref{fig:alertstructure} (right).
As described earlier, the GSR is organized by an
independent third party (MITRE) and the authors of this study do not
have any participation in this activity.

\subsection{Lead Time vs Accuracy of Forecast Date}
\begin{figure}[t]
\centering
\includegraphics[width=0.40\textwidth]{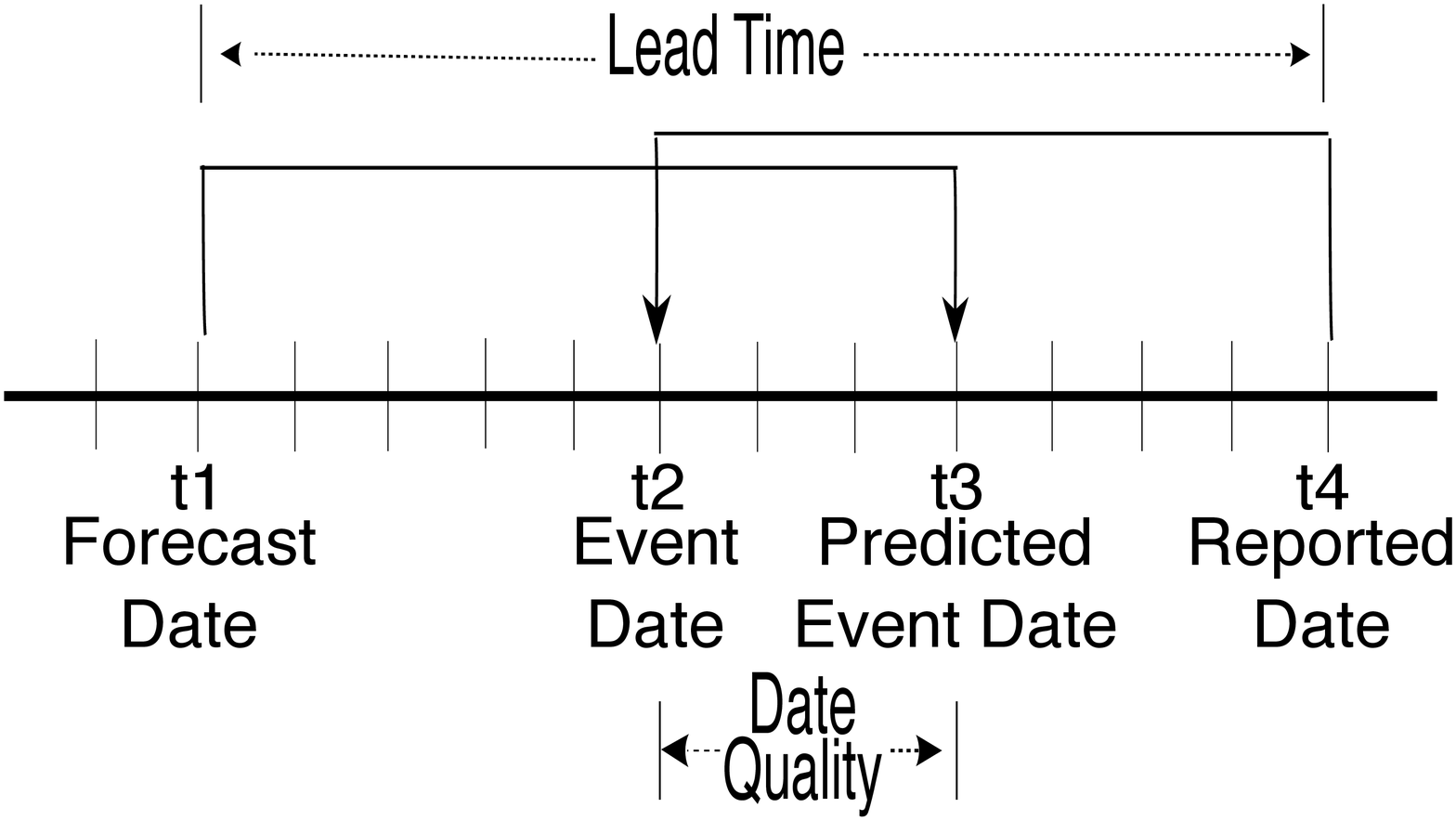}
\caption{Alert sent at time $t1$ predicting an event at time $t3$
can be matched to a GSR event that happened at time $t2$ and reported
at time $t4$ if $t1 < t4$.}
\label{fig:timeline}
\end{figure}

Before we explain how alerts are matched to events, it is important to
first understand which alerts {\it can} be matched to specific events.
Note that there are four dates in an (alert,event) combination (see Fig.~\ref{fig:timeline}):
\begin{enumerate}
\item The date the forecast is made ({\it forecast date})
\item The date the event is predicted to happen ({\it predicted event date})
\item The date the event actually happens ({\it event date})
\item The date the event is reported in a GSR source ({\it reported date})
\end{enumerate}

For an event to be qualified as having been predicted by a warning,
{\it forecast date} $<$ {\it reported date}
(recall that time is measured in
granularities of days).
The {\it lead time} is given as
$(\textit{reported date} - \textit{forecast date})$, i.e., the number
of days by which we `beat the news.'
In contrast, the difference between
{\it predicted event date} and {\it event date}, i.e.,
$|\textit{event date} - \textit{predicted event date}|$.
is one of {\it quality} or accuracy.
Ideally we require {\it lead time} to be as high as possible and
$|\textit{event date} - \textit{predicted event date}|$ to be as low
as possible.

\subsection{Other Quality Aspects}
Forecasting the event date accurately is only one aspect of quality.
Recall that alerts also forecast the location, event type,
and population. We define scores for each of these aspects and
quality is defined as a sum over all these scores.
$$\textrm{Quality score } (QS) = DS + LS + ES + PS$$
where DS, LS, ES, and PS denote the date score, location score,
event type score, and population score, respectively.
Each of these scores is in turn defined next:
$$DS = 1 - \min(|\textit{event date} - \textit{predicted event date}|,7)/7$$
If the date of the event listed in the warning is the same as the
actual date of the event, then $DS$ is 1. On the other hand, if these dates
are farther than 7 days apart, then $DS$ is 0.

Location score (LS) can be defined in many ways. Because location is
defined in terms of triples of (country, state, city), one approach is
to use a tiered formula. Comparing a GSR event with a warning, we can
obtain a score triple of $(l_1, l_2, l_3)$ where $l_1$ is the
country-level score, $l_2$ is the state-level score, and $l_3$ is the
city-level score. Each of these scores have a value of $0$ if they
do not match and $1$ is they match. Then the match between submitted
warning location and the GSR location is given by:
$$LS = {1 \over 3} l_1 + {1 \over 3} l_1 l_2 + {1 \over 3} l_1 l_2 l_3$$
\noindent
An alternative way to define location score is as:
$$LS = 0.33 + 0.66 (1 - \min(\textrm{dist},300)/300)$$
where $\textrm{dist}$ denotes the distance (in km) between the city predicted and
the GSR city. All city location names are standardized to the World Gazetteer which provides
latitude and longitude values, thus facilitating the computation of distance.
The scaling and shifting values of 0.33 and 0.66 ensure that this definition of LS
is compatible with the earlier definition. Cities outside a 300km radius from a GSR location
will thus be scored 0.33; exact predictions will be scored 1; and cities within a 300km radius
will get scores in the range [0.33,1].
We distinguish between these two criteria as the categorical LS versus
physical distance-based LS.

Event type score (ES) is scored similar to categorical LS since it naturally maps to
a three-level taxonomy: whether a civil unrest is forecast to happen,
what objective/reason is behind the unrest, and whether it is violent.
Again partial credit applies depending on the level of specification.
Population score (PS) is simply a binary (0/1) score denoting whether we
forecast the correct population or not.
Finally, note that $QS = DS + LS + ES + PS$ is designed to take values in the
range $[0,4]$.

\subsection{Inclusion Criteria}
Thus far we have demonstrated, given a warning-event pair, how we can
score their fitness. Inclusion criteria define which W-E pairs {\it can}
even be considered for scoring. We have already mentioned one inclusion
criterion, viz. that lead time must be $> 0$. The full list of inclusion
criteria we will consider are:
\begin{enumerate}
\item Lead time $> 0$
\item Both warning and event are for the same country.
\item The {\it predicted event date} and {\it event date} must be
within 7 days of each other.
\end{enumerate}
A fourth, optional (and stringent), criterion we will use is:
\begin{enumerate}
  \setcounter{enumi}{3}
  \item Both predicted location and event location must be within 300km of
each other.
\end{enumerate}
It is important to distinguish the inclusion criteria from the scoring
criteria. Inclusion criteria define which W-E pairs are allowable.
Scoring criteria determine, from these allowable W-E pairs, what their
score will be.

\subsection{Matching Alerts to Events}
\begin{figure}[t]
\centering
\includegraphics[width=0.50\textwidth]{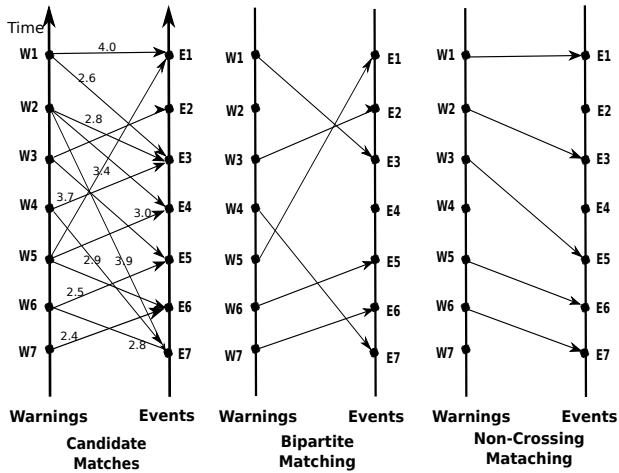}
\caption{Given a set of candidate warning-event matches (left), we evaluate the
performance of EMBERS using either a regular bipartite matching (middle) or
by constructing a non-crossing matching (right).}
\label{fig:matching}
\end{figure}

Thus far we have assumed that we are matching an alert to a GSR event. In
practice, the problem is we are given a set of issued alerts  and a set
of GSR events and we must determine the quality of the match: which
alert would correspond to which event? One strategy is to construct
a bipartite graph between the set of alerts and the set of events,
where allowable edges are those that satisfy the inclusion criteria, and
where weights on these allowable edges denote their quality scores.
We then construct
a maximum weighted bipartite matching, e.g., see
Fig.~\ref{fig:matching} (middle). Such matchings are conducted on a monthly basis
with a lookback period to bring in unmatched warnings from the previous month.

\subsection{Non Crossing Matching}
A criticism of the matching approach above is that it can lead to
criss-cross matches, i.e., the matching process may not respect the
temporal order in which warnings were issued or in which events unfold.
A non-crossing matching is a more restrictive version of a bipartite
matching. Consider two warnings $w_1$ and $w_2$ and two events $e_1$ and
$e_2$. Representing them by their predicted event dates and
event dates, and assuming $w_1 < w_2$ and $e_1 < e_2$, then
$\{(w_1, e_2),(w_2, e_1)\}$ is a {\it crossing matching} since the
earlier warning is paired to a later event (and vice versa). To respect
the temporal order, we also investigate the computation of a maximum
non-crossing matching~\cite{non-cross-cite} and use it as
an additional evaluation criterion (see Fig.~\ref{fig:matching} (right)).

\subsection{Putting it all together: Five criteria}
We are now ready to identify all the evaluation criteria used in EMBERS.
The overall {\bf quality} is defined as a weighted average
across all matched warning-event pairs.
Similarly, {\bf lead time} is averaged across all matched warning-event pairs.
In addition, we can define {\bf precision} in terms of the number of unmatched
warnings as a fraction of the total number of sent warnings.
Similarly, {\bf recall} can be defined in terms of the number of unmatched
events as a fraction of the total number of events.
Finally, a {\bf probability score} is calculated over all warnings,
mapped or unmapped. For each warning,
it is defined in terms of the Brier score, i.e.,
$1 - (o-p)^2$ where
$p$ is the probability assigned to the warning, and $o$ is 1 if
the warning is mapped to some event in the GSR, and 0 is the warning
is not mapped to an event in the GSR.  This score is then averaged over
all warnings.

\vspace{-1em}
\section{Evaluation Results}
\begin{figure*}[t]
  \centering
  \begin{subfigure}[b]{0.32\textwidth}
      \includegraphics[width=\textwidth]{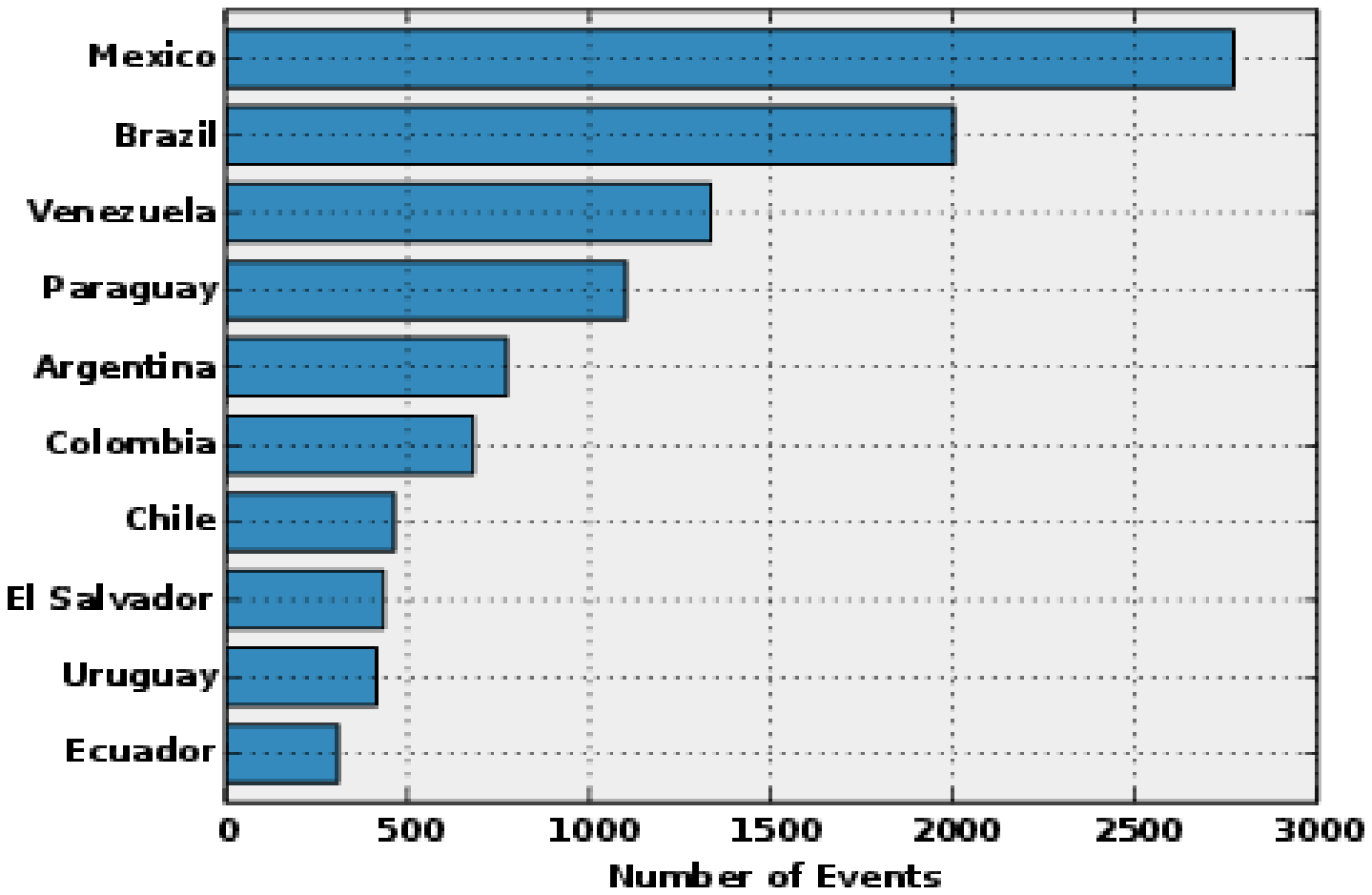}
      \caption{GSR Distribution}
      \label{fig:gsrdistribution}
  \end{subfigure}
  \begin{subfigure}[b]{.32\textwidth}
    \includegraphics[width=\textwidth]{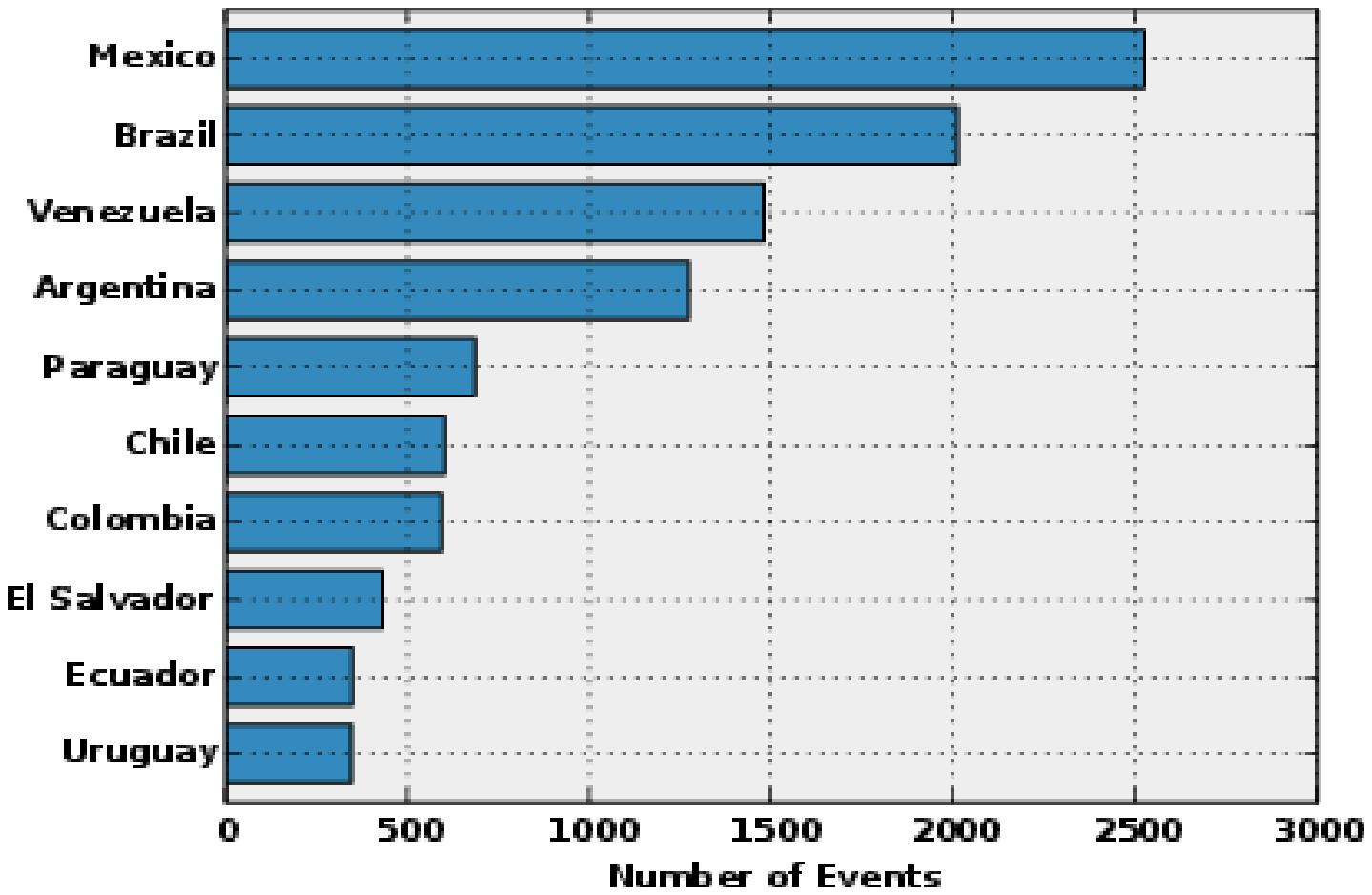}
    \caption{EMBERS Alerts Distribution}
    \label{fig:embersdistribution}
  \end{subfigure}
  \begin{subfigure}[b]{.32\textwidth}
    \includegraphics[width=\textwidth]{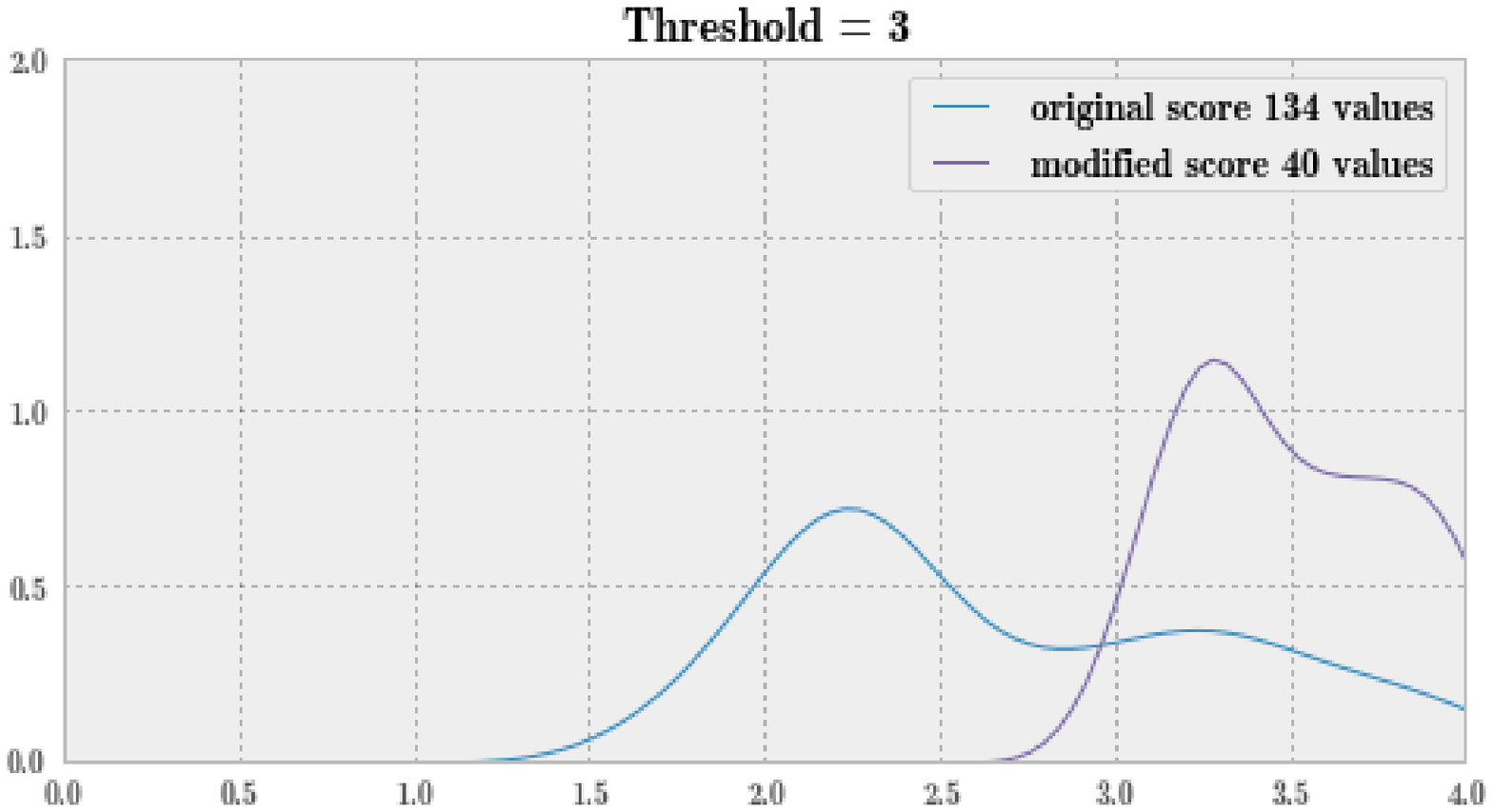}
    \caption{Quality distributions before and after suppression/}
    \label{fig:supressor}
  \end{subfigure}
  \begin{subfigure}[b]{.32\textwidth}
    \centering
    \includegraphics[width=\textwidth, height=0.2\textheight]{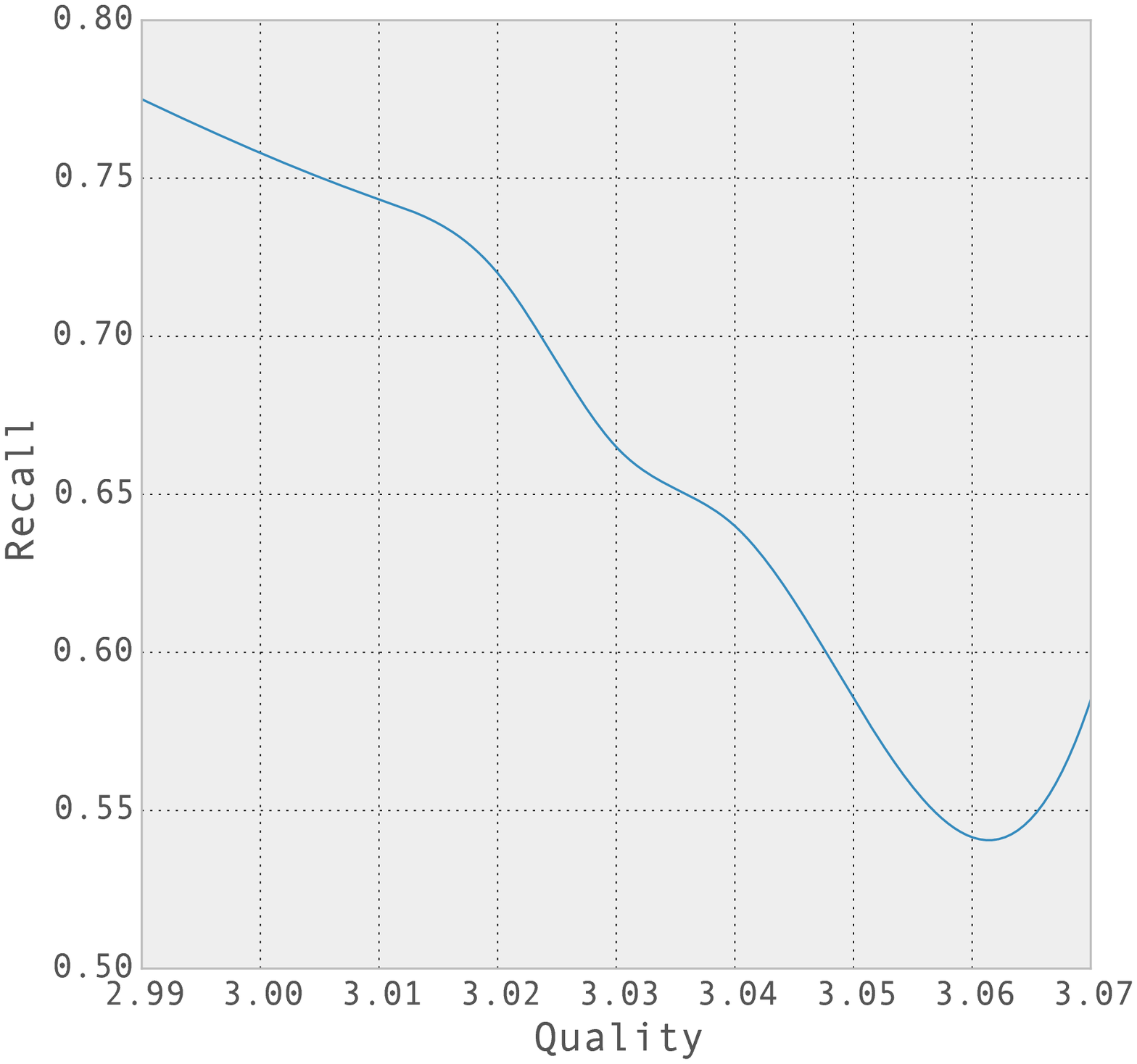}
    \caption{Recall vs Quality Tradeoff}
    \label{fig:recallQuality}
  \end{subfigure}
  \begin{subfigure}[b]{.32\textwidth}
    \includegraphics[width=\textwidth]{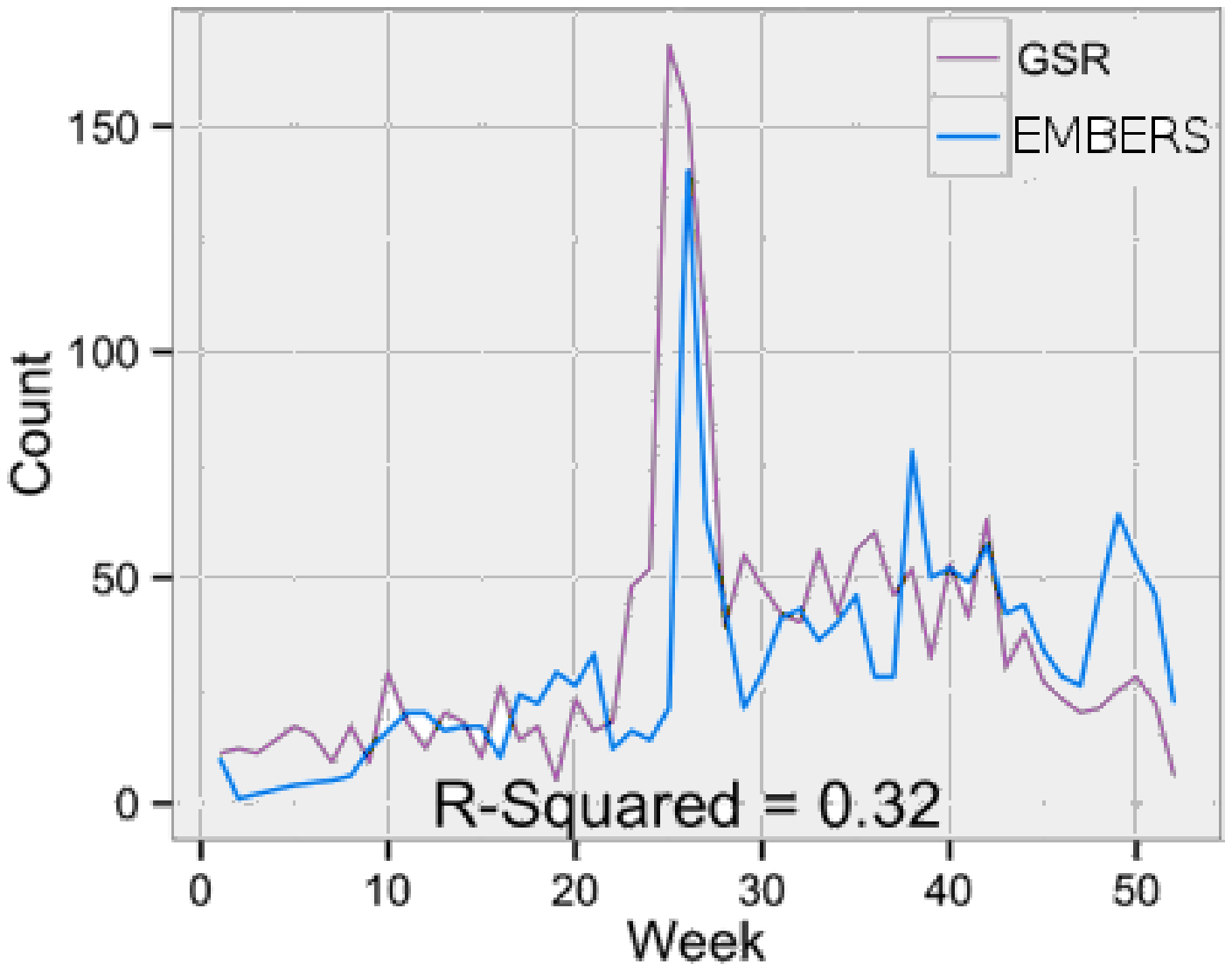}
    \caption{The Brazilian Spring}
    \label{fig:brazil_june}
  \end{subfigure}
  \begin{subfigure}[b]{.32\textwidth}
    \includegraphics[width=\textwidth]{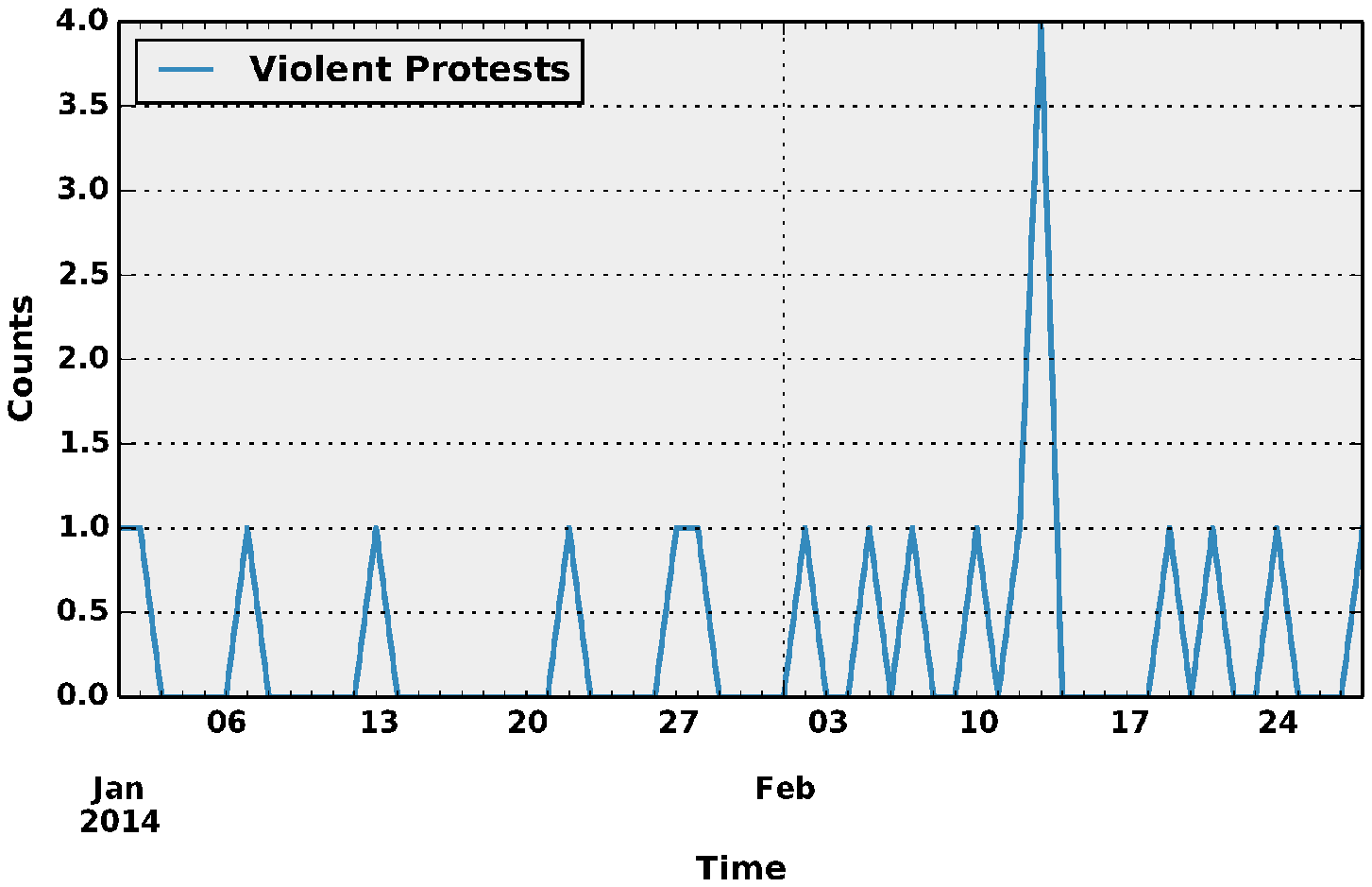}
    \caption{Violent Protests in Brazil}
    \label{fig:violent_brazil}
  \end{subfigure}
  \begin{subfigure}[b]{.32\textwidth}
    \includegraphics[width=\textwidth]{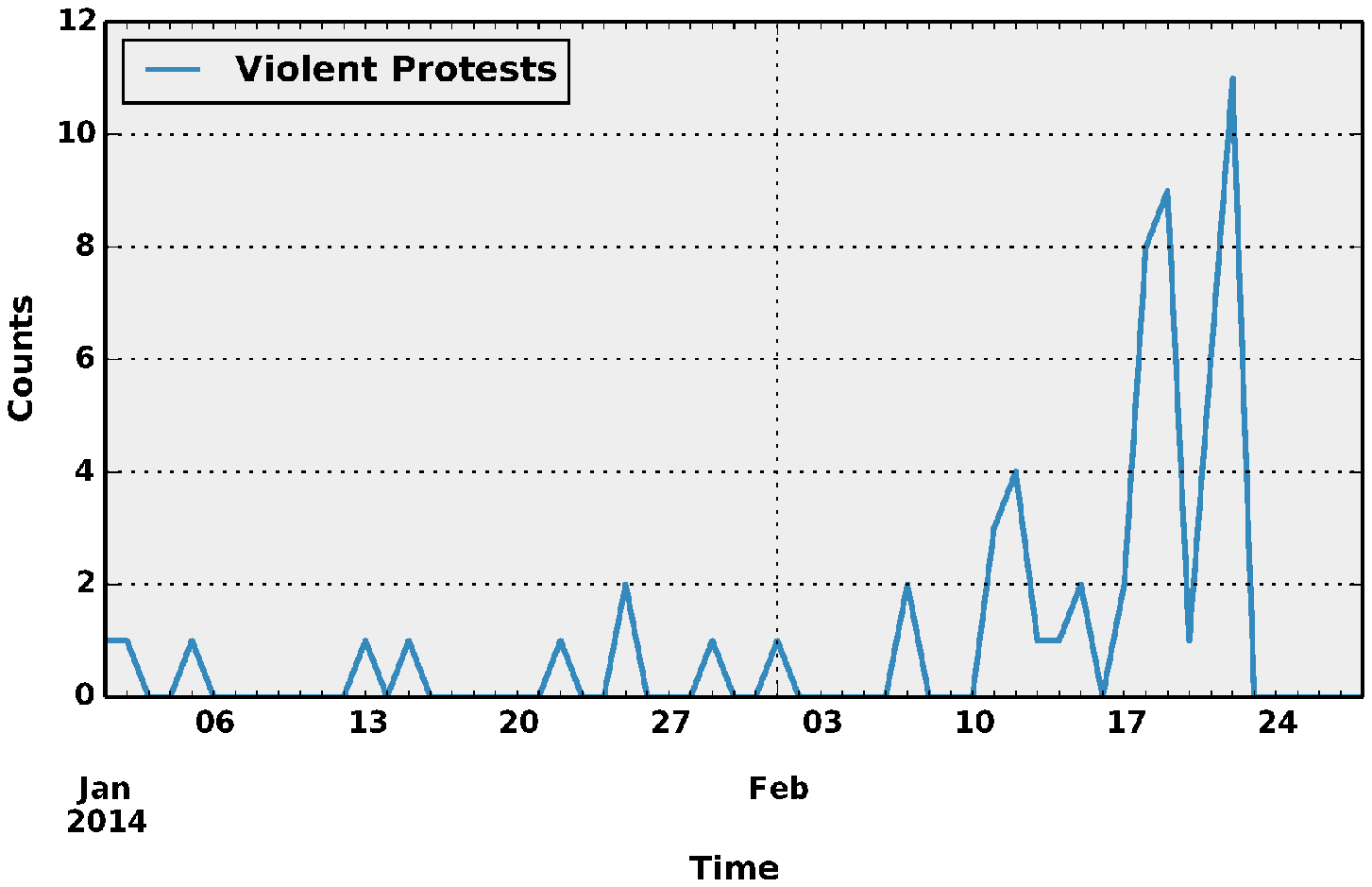}
    \caption{Violent Protests in Venezuela}
    \label{fig:violent_venezuela}
  \end{subfigure}
  \begin{subfigure}[b]{0.32\textwidth}
      \includegraphics[width=\textwidth]{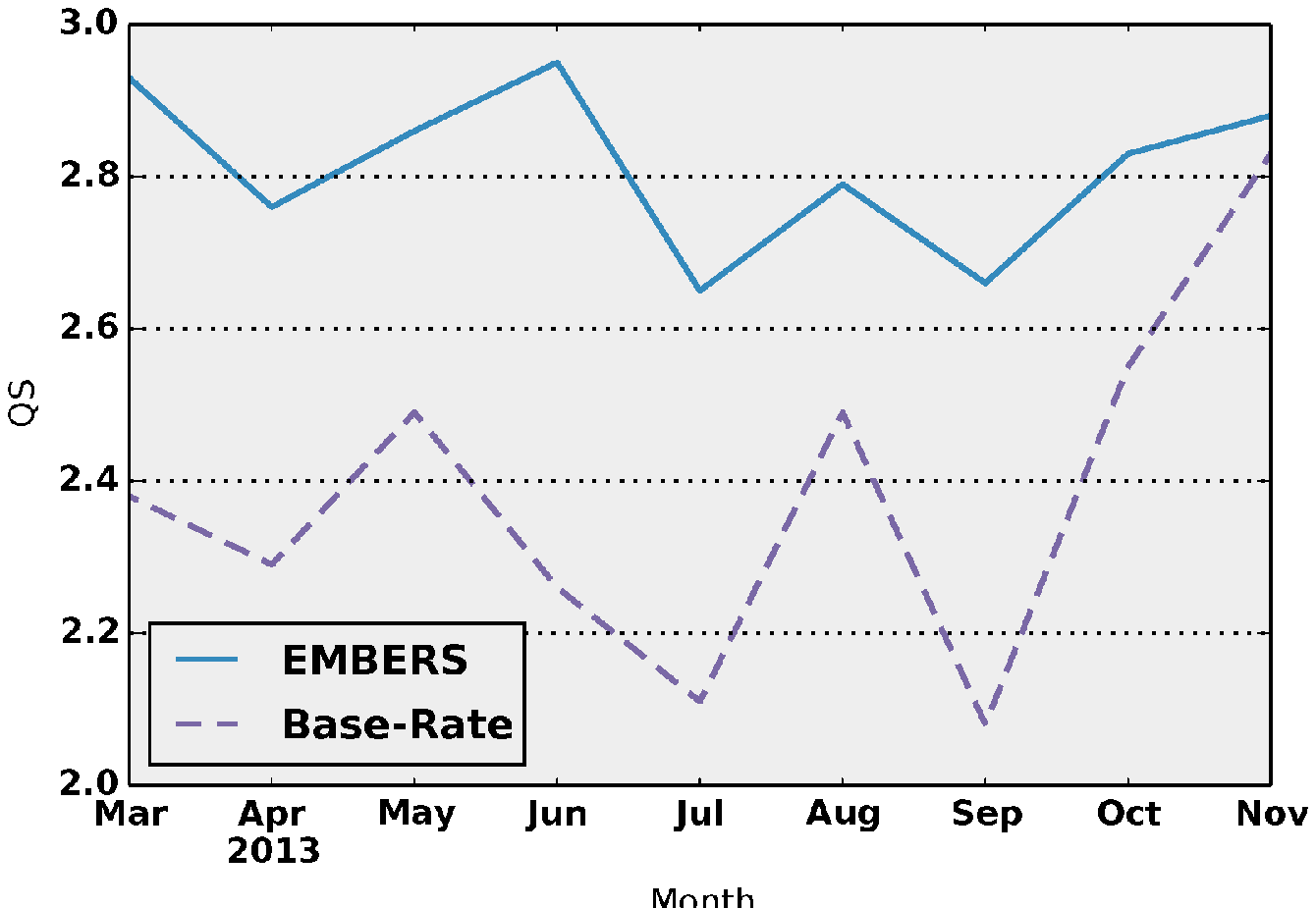}
      \caption{MaxEnt Evaluation}
      \label{fig:maxent}
  \end{subfigure}
  \begin{subfigure}[b]{.32\textwidth}
    \includegraphics[width=\textwidth]{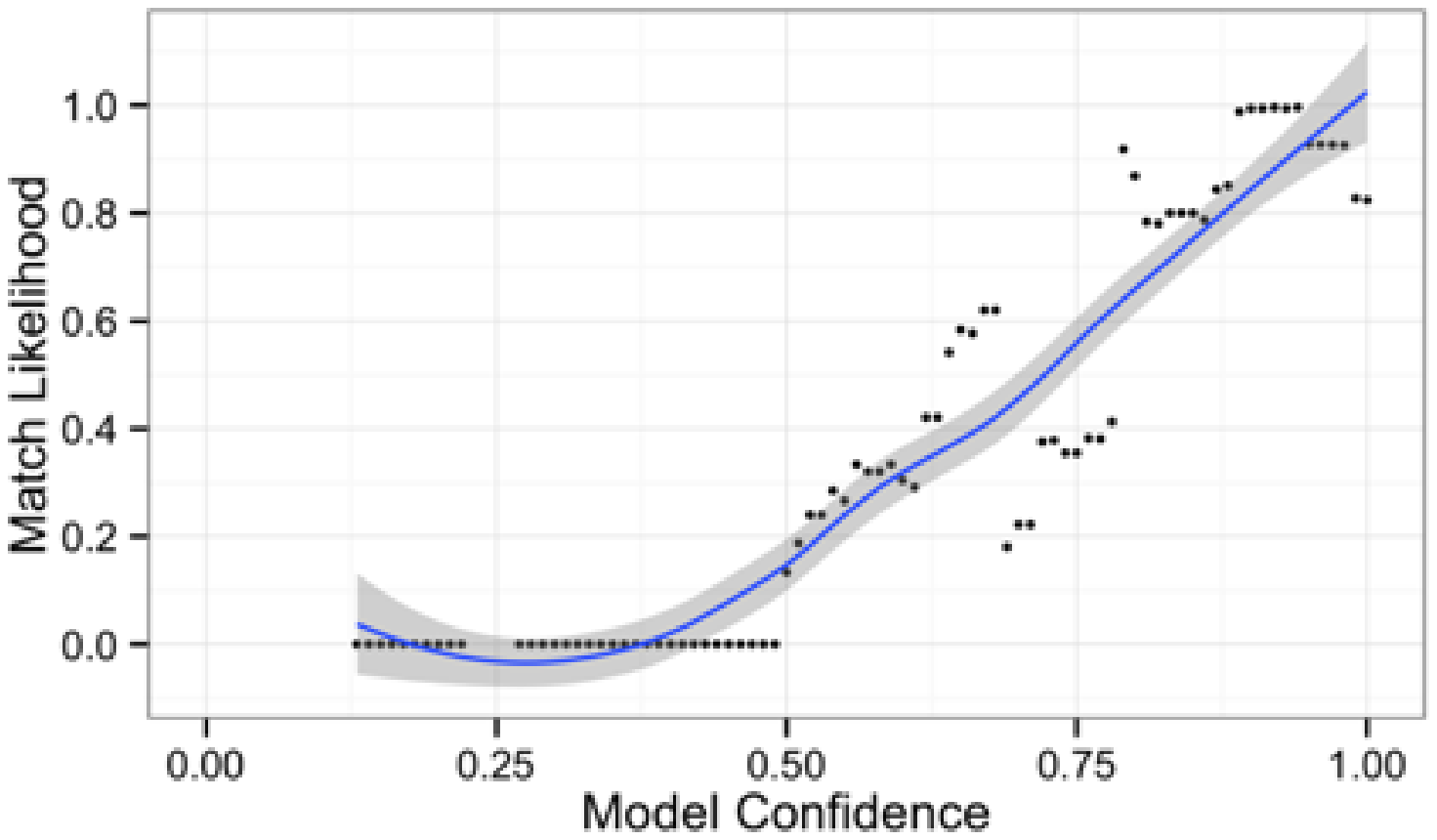}
    \caption{EMBERS Probability Score}
    \label{fig:probscore}
  \end{subfigure}
  \begin{subfigure}[b]{.32\textwidth}
    \includegraphics[width=\textwidth,height=0.7\textwidth]{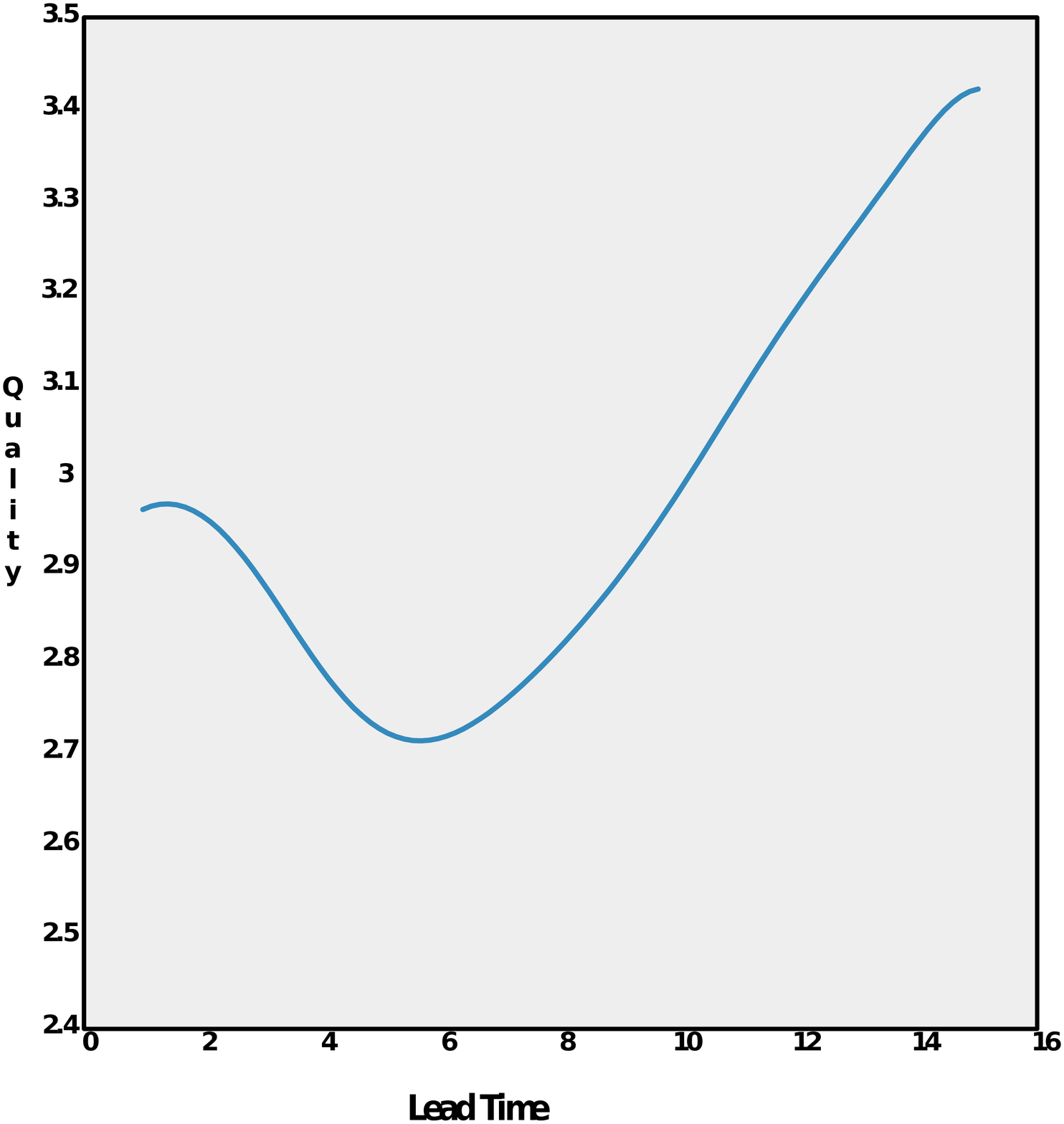}
    \caption{Lead time vs Quality Tradeoff}
    \label{fig:leadvsqs}
  \end{subfigure}
  \begin{subfigure}[b]{.32\textwidth}
    \includegraphics[width=\textwidth]{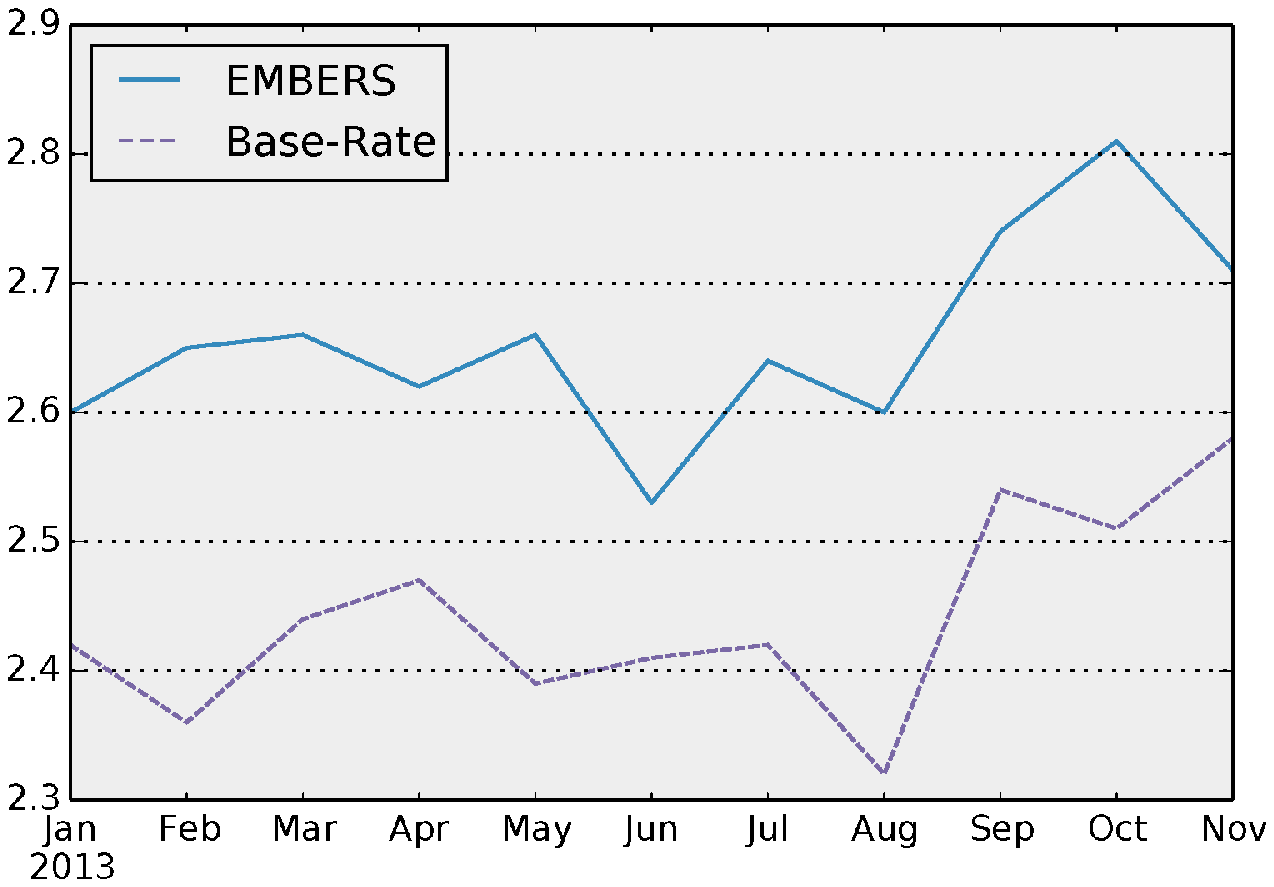}
    \caption{Quality score with non-crossing matches (1 day interval).}
    \label{fig:noncrossing}
  \end{subfigure}
  \begin{subfigure}[b]{.32\textwidth}
    \centering
    \includegraphics[width=\textwidth]{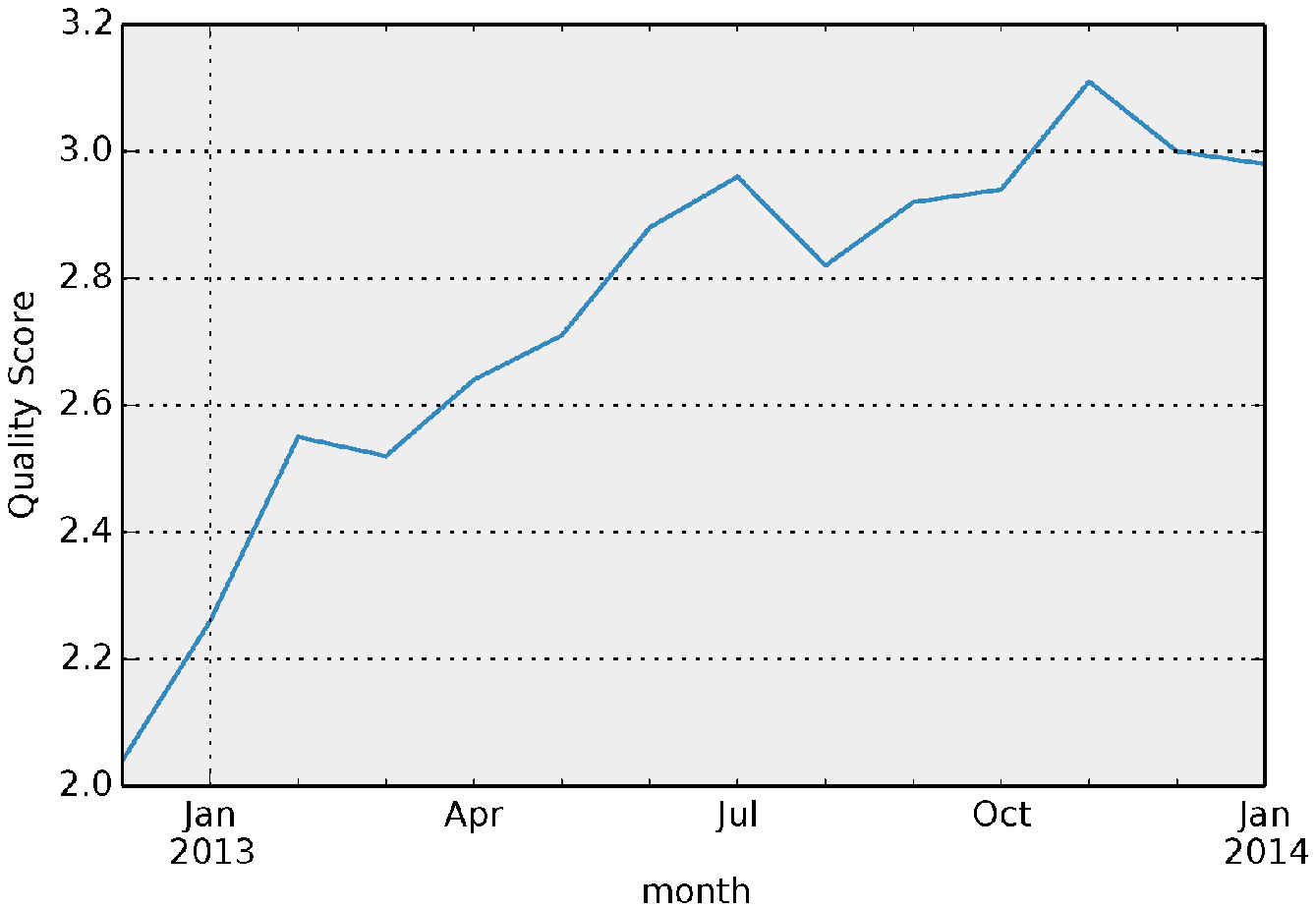}
    \caption{Evolution of EMBERS's quality scores over time.}
    \label{fig:monthlyqs}
  \end{subfigure}
  \caption{Evaluation of EMBERS}
\end{figure*}

We present an exhaustive evaluation of EMBERS against multiple aspects as follows:

\begin{itemize}
\item {\bf How do each of our models fare for the 10 countries of interest and how well does their integration
achieve the five overall metrics?}
\end{itemize}
Table~\ref{tb:modelwisecomparison} presents the performance of EMBERS models for a recent month across
all the 10 countries of interest here. As is clear here, the models have selective superiorities
across the countries studied. While the baseline model captures significant regularities and achieves
high quality scores, models like DQE perform better for countries like Brazil, Mexico, and Venezuela,
all of which have significant chatter on Twitter. Even when models have comparable performances, their
integration is useful because each model will produce only a limited set of warnings and their fusion is
necessary to achieve high recall. This is evident in Table~\ref{tb:pmmetrics} that demonstrates that
we achieve a quality score of 3.11 with an average lead time of 8.8 days and respectable precision and
recall (0.69 and 0.82, respectively). Taking a birds eye view,
Fig.~\ref{fig:gsrdistribution} and Fig.~\ref{fig:embersdistribution}
summarize the distribution of events in the GSR and
alerts sent by EMBERS over the past 15 months.

\begin{itemize}
\item {\bf How does EMBERS's fusion and suppression engine help `shape' our quality distribution?}
\end{itemize}
Fig.~\ref{fig:supressor} describe how our suppression engine can be tuned to steer the quality distribution from a mode
around 2.25 to one around 3.2 by learning which warnings to suppress and which ones to issue. This capability directly
helps balance the recall-quality tradeoff, as shown in
Fig.~\ref{fig:recallQuality}.  %

\begin{itemize}
\item {\bf How does EMBERS fare against a baserate model with lenient versus stringent inclusion criteria
for matching?}
\end{itemize}
To rigorously evaluate
the capabilities of EMBERS, we implemented a baserate model as a yardstick for comparison. The baserate model
is similar in spirit to the baseline model described earlier but functions differently. Rather than
filtering for frequent combinations of event properties, it generates alerts using the rate of occurrence of events in the past
three months. Due to space considerations we are unable to describe these results in detail. Under the lenient (categorical) inclusion
criteria constraints, EMBERS exhibits a quality score improvement of approximately +0.4 over baserate methods. Under the
strict location-based inclusion criteria, this improvement jumps to a +1.0 over baserate methods.

\begin{itemize}
\item {\bf How adept is EMBERS at forecasting `surprising' events?
Did EMBERS forecast significant uprisings such as the June 2013 protests in Brazil?}
\end{itemize}
Fig.~\ref{fig:brazil_june} describe the performance of our system in Brazil during the summer of 2013 when Brazil witnessed
significant protests that were originally triggered by bus fare increases. As can be seen, EMBERS is able to track the rise in
number of protests quite accurately. More recently,
Fig.~\ref{fig:violent_brazil} and Fig.~\ref{fig:violent_venezuela} describe the performance of
EMBERS in Brazil and Venezuela for the Jan-Feb 2014 season. Significant violent protests were witnessed in both countries,
due to bus fare increases and student-led demonstrations, respectively. While the GSR for Feb 2014 is not available at the
time of this writing, it is clear that the uptick in violence in both countries is captured in EMBERS alerts forecasting
violence. Finally, we also conducted a formal maximum entropy evaluation of protest counts, to determine how EMBERS fares
on only those protests that are deemed to significantly higher in number relative to the past three months. As
Fig.~\ref{fig:maxent} %
shows, EMBERS demonstrates an improvement of nearly 0.5 over baserate models during months of significant uprisings (e.g., June
2013). During other months (e.g., Nov 2013) there is relatively normal activity and baserate methods perform comparably.

\begin{itemize}
\item {\bf How reliable are EMBERS's probability scores?}
\end{itemize}
Fig.~\ref{fig:probscore} shows that the probability scores emitted by warnings have a monotonic relationship to the likelihood of
matches, indicating that EMBERS's use of confidence captures the mapping from model and warning attributes to the
possibility of event matches.

\begin{itemize}
\item {\bf How does EMBERS's lead time vary with quality scores?}
\end{itemize}
Fig.~\ref{fig:leadvsqs} illustrates an interesting relationship. As lead time increases from low values, as expected,
quality scores decrease. But as lead time crosses a threshold, quality scores actually improve again! This is because
data sources like Facebook event pages and other feeds contribute high quality planned protest warnings with high lead time.

\begin{itemize}
\item {\bf What is the effect of adopting regular versus non-crossing matching constraints?}
\end{itemize}
Fig.~\ref{fig:noncrossing} reveals that, as expected, when adopting non-crossing matching constraints, the number of matches
decreases bringing down the overall quality. Nevertheless, a consistent level of improvement over baserate methods is witnessed.

\begin{itemize}
\item {\bf How has the performance of EMBERS improved over time?}
\end{itemize}
Finally, Fig.~\ref{fig:monthlyqs} demonstrates the performance of our deployed EMBERS system over time. From quality scores of just
over 2 in the past year, EMBERS has breached the 3.0 barrier in recent months.

\begin{table*}[tb!]
\centering
\caption{Comparing the forecasting accuracy of
different models in EMBERS. Quality scores in this and other tables are in the range [0,4]
where 4 is the most accurate. AR=Argentina; BR=Brazil; CL=Chile; CO=Colombia; EC=Ecuador;
SV=El Salvador; MX=Mexico; PY=Paraguay; UY=Uruguay; VE=Venezuela. A $--$ indicates that
the model did not produce any warnings for that country in the studied period.}
\label{tb:modelwisecomparison}
\begin{tabular}{||l|*{17}{c|}}
\hline
Model & AR & BR & CL & CO & EC & SV & MX & PY & UY & VE & All\\
\hline
Dynamic Query Expansion &3.1&3.31&1.88&3.1&2.43&2.94&3.26&2.88&2.72&2.9&2.97\\
\hline
Volume-based Model &3.0&3.11&-&2.9&-&-&3.15&-&1.72&2.9&2.88\\
\hline
MLE   &3.33&3.0&2.87&3.15&2.29&3.11&3.11&3.1&2.57&2.77&3.0\\
\hline
Planned Protest &2.59&2.64&2.4&2.85&1.92&-&3.0&2.89&2.85&2.66&2.76\\
\hline
Cascades Model &3.13&-&-&-&-&-&-&-&-&2.93&3.0\\
\hline
\end{tabular}
\end{table*}

\begin{table*}
\centering
\caption{EMBERS metrics across multiple countries. }
\label{tb:pmmetrics}
\begin{tabular}{||l|*{17}{c|}}
\hline
Metric & AR & BR & CL & CO & EC & MX & PY & SV & UY & VE & All\\
\hline
Quality score &3.2&3.39&2.85&2.86&2.59&3.0&3.27&2.85&3.05&3.01&3.11\\
\hline
Recall &1.0&1.0&0.82&0.59&1.0&1.0&0.65&1.0&1.0&0.84&0.82\\
\hline
Precision &0.55&0.45&0.89&0.94&0.77&0.71&1.0&0.69&0.46&0.73&0.69\\
\hline
Lead time (days) &10.44&11.82&6.25&7.85&8.44&8.32&8.61&10.57&8.8&6.03&8.88\\
\hline
Probability measure &0.71&0.66&0.87&0.87&0.75&0.74&0.94&0.74&0.72&0.72&0.76\\
\hline
\end{tabular}
\end{table*}

\vspace{-1em}
\section{Discussion}
We have presented the architecture of EMBERS, an automated system
for generating forecasts about civil unrest from massive, multiple, data sources.
Our evaluations over 10 countries illustrate the capabilities of EMBERS `in the small' (matching
specific events to particular warnings) as well as `in the
large' (capturing significant upticks across countries). 

Future work is targeted at three aspects. First, we are interested in social science theory-based
approaches to forecasting, e.g., modeling the rise of grievances into trigger events, capturing the
role of opinion leaders, and identifying whether there are both necessary and sufficient conditions
for a festering sentiment to transform into a protest. Second, we plan to develop a statistical theory
of tradeoffs revolving around the boundaries of precision-recall and quality-lead time. Different analysts
are likely to prefer different sweet spots along these boundaries and we seek to situate EMBERS as a tunable
forecasting system.
Finally, for analyst consumption, we are interested in automated narrative generation, i.e., an English description of
an alert providing a contextual summary of the alert (similar to automated weather report generation).

\vspace{-1em}
\section*{Acknowledgments}
{\small Supported by the Intelligence Advanced Research Projects Activity (IARPA) via
DoI/NBC contract number D12PC000337, the US Government is authorized to reproduce and distribute reprints of
this work for Governmental purposes notwithstanding any copyright annotation thereon.
Disclaimer: The views and conclusions contained herein are those of the authors and should not be interpreted as necessarily representing the official policies or endorsements, either expressed or implied, of IARPA, DoI/NBC, or the US Government.

\clearpage
\bibliographystyle{abbrv}
\bibliography{references}

\begin{thebibliography}{10}

\bibitem{gravano}
H.~Becker, D.~Iter, M.~Naaman, and L.~Gravano.
\newblock {Identifying content for planned events across social media sites}.
\newblock In {\em WSDM 2012}. ACM Press, 2012.

\bibitem{Box-Steffensmeier2004}
J.~M. Box-Steffensmeier and B.~S. Jones.
\newblock {\em {Event History Modeling: A Guide for Social Scientists}}.
\newblock 2004.

\bibitem{braha}
D.~Braha.
\newblock {Global civil unrest: contagion, self-organization, and prediction.}
\newblock {\em PLoS One}, 7(10):e48596, Jan. 2012.

\bibitem{texasguy}
P.~T. Brandt, J.~R. Freeman, and P.~A. Schrodt.
\newblock Real time, time series forecasting of inter-and intra-state political
  conflict.
\newblock {\em Conflict Management and Peace Science}, 28(1):41--64, 2011.

\bibitem{broecheler:uai10}
M.~Broecheler, L.~Mihalkova, and L.~Getoor.
\newblock Probabilistic similarity logic.
\newblock In {\em Proc. UAI}, 2010.

\bibitem{sandra}
S.~Gonz\'{a}lez-Bail\'{o}n, J.~Borge-Holthoefer, A.~Rivero, and Y.~Moreno.
\newblock {The dynamics of protest recruitment through an online network.}
\newblock {\em Sci. Rep.}, 1:197, Jan. 2011.

\bibitem{tingpaper}
T.~Hua, C.-T. Lu, N.~Ramakrishnan, F.~Chen, J.~Arredondo, D.~Mares, and
  K.~Summers.
\newblock {Analyzing Civil Unrest through Social Media}.
\newblock {\em Computer}, 46(12):80--84, Dec. 2013.

\bibitem{newpaperbyJason}
N.~Kallus.
\newblock {Predicting Crowd Behavior with Big Public Data}.
\newblock Feb. 2014.

\bibitem{Leetaru2013}
K.~Leetaru and P.~Schrodt.
\newblock {GDELT: Global data on events, location, and tone, 1979–2012}.
\newblock {\em ISA Annual Convention}, pages 1979--2012, 2013.

\bibitem{Leetaru2011}
K.~H. Leetaru.
\newblock {Culturomics 2.0: Forecasting large-scale human behavior using global
  news media tone in time and space}.
\newblock {\em First Monday}, 16:1--22, 2011.

\bibitem{LlorensDGS12}
H.~Llorens, L.~Derczynski, R.~J. Gaizauskas, and E.~Saquete.
\newblock {\sc TIMEN}: An open temporal expression normalisation resource.
\newblock In {\em LREC}, pages 3044--3051. European Language Resources
  Association, 2012.

\bibitem{non-cross-cite}
F.~Malucelli, T.~Ottmann, and D.~Pretolani.
\newblock Efficient labelling algorithms for the maximum noncrossing matching
  problem.
\newblock {\em {Discrete Applied Mathematics}}, 47(2):175--179, 1993.

\bibitem{icews}
S.~P. O’Brien.
\newblock {Crisis Early Warning and Decision Support: Contemporary Approaches
  and Thoughts on Future Research}.
\newblock {\em Int. Stud. Rev.}, 12(1):87--104, Mar. 2010.

\bibitem{etzioni}
A.~Ritter, O.~Etzioni, and S.~Clark.
\newblock {Open domain event extraction from twitter}.
\newblock In {\em ACM SIGKDD 2012}, 2012.

\bibitem{philschrodt}
P.~A. Schrodt.
\newblock Automated production of high-volume, near-real-time political event
  data.
\newblock In {\em American Political Science Association meetings}, 2010.

\bibitem{Tibshirani1996}
R.~Tibshirani.
\newblock {Regression shrinkage and selection via the lasso}.
\newblock {\em J. R. Stat. Soc. Ser. B}, 58:267--288, 1996.

\end{thebibliography}
\end{document}